# Mesophase Formation Stabilizes High-purity Magic-sized Clusters

Douglas R. Nevers†§, Curtis B. Williamson†§, Benjamin H. Savitzky⁺, Ido Hadar#, Uri Banin#, Lena F. Kourkoutis∥,⊥, Tobias Hanrath†*, and Richard D. Robinson‡*

†Robert F. Smith School of Chemical and Biomolecular Engineering, Cornell University, Ithaca, USA. ⁺Department of Physics, Cornell University, Ithaca, USA. #The Institute of Chemistry and Center for Nanoscience and Nanotechnology, The Hebrew University of Jerusalem, Jerusalem 91904, Israel ∥School of Applied and Engineering Physics, Cornell University, Ithaca, USA. ⊥Kavli Institute for Nanoscale Science, Cornell University, Ithaca, USA. ‡Department of Materials Science and Engineering, Cornell University, Ithaca, USA.

**ABSTRACT:** Magic-sized clusters (MSCs) are renowned for their identical size and closed-shell stability that inhibit conventional nanoparticle (NP) growth processes. Though MSCs have been of increasing interest, understanding the reaction pathways toward their nucleation and stabilization is an outstanding issue. In this work, we demonstrate that high concentration synthesis (1000 mM) promotes a well-defined reaction pathway to form high-purity MSCs (>99.9%). The MSCs are resistant to typical growth and dissolution processes. Based on insights from *in-situ* X-ray scattering analysis, we attribute this stability to the accompanying production of a large, hexagonal organic-inorganic mesophase (>100 nm grain size) that arrests growth of the MSCs and prevents NP growth. At intermediate concentrations (500 mM), the MSC mesophase forms, but is unstable, resulting in NP growth at the expense of the assemblies. These results provide an alternate explanation for the high stability of MSCs. Whereas the conventional mantra has been that the stability of MSCs derives from the precise arrangement of the inorganic structures (*i.e.*, closed-shell atomic packing), we demonstrate that anisotropic clusters can also be stabilized by self-forming fibrous mesophase assemblies. At lower concentration (<200 mM or >16 acid-to-metal), MSCs are further destabilized and NPs formation dominates that of MSCs. Overall, the high concentration approach intensifies and showcases inherent concentration-dependent surfactant phase behavior that is not accessible in conventional (*i.e.,* dilute) conditions. This work provides not only a robust method to synthesize, stabilize, and study identical MSC products, but also uncovers an underappreciated stabilizing interaction between surfactants and clusters.

## Introduction

Traditionally, colloidal nanoparticle (NP) synthesis is characterized by solely tracking the evolution of the inorganic materials from precursor conversion to monomers and, ultimately, to NPs.[1–4] Recent studies demonstrate that the organic surfactants play a central role during NP synthesis by controlling the precursor solubility and reactivity.[1,2,5–8] In addition to these critical functions, the NP cation precursors alone, as an isolated system, also exhibit well-established surfactant phase behavior, even at elevated temperatures (>100°C), and were previously known as heavy metal soaps (*i.e.*, metal carboxylates).[6,9,10] Only recently has the surfactant behavior of NP precursors become appreciated within the NP field.[6,11–13]

For instance, Buhro and co-workers reported that, at lower temperatures, MSCs, which are single sized nanomaterial, form within a lamellar surfactant mesophase or liquid crystalline structure that is composed of the precursors.[11,12,14] Mesophases are partially ordered structures (*e.g.*, liquid crystals) and are characterized by a turbid solution,[15] well-defined peaks in small-angle X-ray scattering (SAXS),[16,17] and/or optical birefringence.[18] Based on these metrics, several studies have alluded to a connection between surfactant structure and MSC formation such as the observation of solution turbidity and the self-assembly of MSCs,[19] or detection of large (~1 nm) micellar aggregates.[20,21] These results suggest that previous studies may have been, unknowingly, observing surfactant phase behavior, and in some cases mesophase formation, coupled with MSC formation.

MSCs are generally suspected to form in syntheses with higher levels of monomer supersaturation, when precursor conversion kinetics are faster than the nucleation rate, and to function as a reservoir for monomer.[22–27] Previous studies have achieved high levels of supersaturation, and thus promoted MSCs formation, through lower synthesis temperatures (in some cases, <100°C),[11,12,14,21,22,28,29] and low acid-to-metal ratios (~3).[19,25,30–32] If temperatures are sufficiently low (or concentration sufficiently high), monomeric surfactants can assemble into micelles or mesophases based on micelle theory (*i.e.,* above the critical micelle temperature/concentration). This behavior is analogous to monomer nucleation into NPs at sufficiently high supersaturation (*i.e.*, critical nucleation temperature/concentration). Hence, the high supersaturation (of both surfactants and inorganic species) during MSC formation may relate to surfactant phase behavior. While lower temperatures have been directly investigated to achieve higher supersaturation and thus promote MSC formation,[11,12,14,21,28] the importance of high precursor (and thus metal surfactant) concentrations and its relationship to surfactant mesophases have not yet been established.

## Conventional

Magic-sized Cluster 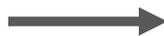 Nanoparticle

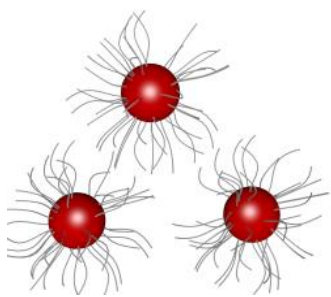 Growth 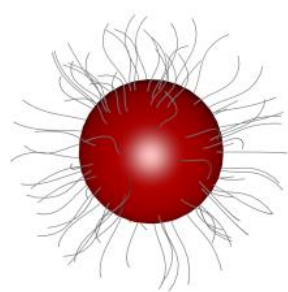

## High Concentration

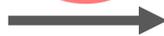 Magic-sized Cluster Assembly

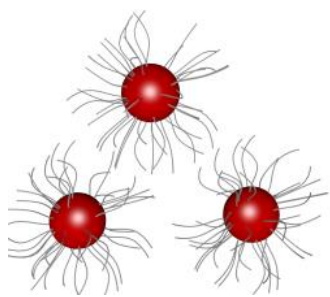 Growth / Assembly 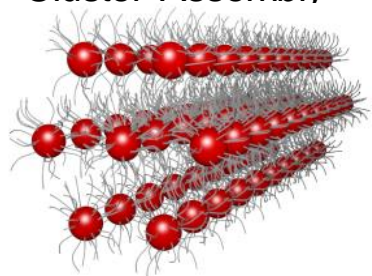

**Figure 1:** *Synthesis Pathways.* (a) Schematic illustrating the fundamental differences in reaction pathways between conventional (100 mM) synthesis and high concentration (1000 mM) synthesis. For conventional synthesis, nucleation and growth occur simultaneously; in contrast, at high concentrations, the synthesis stops after the MSC formation/nucleation because of the formation of a MSC assembly.

In this paper we address the outstanding question: *how does the precursor concentration direct the synthetic pathway between NPs and MSCs?* We show that high precursor (or metal surfactant) concentrations preferentially promote MSCs formation and suppress NP growth. We attribute the suppression of NP growth to the formation of fibrous mesophase assemblies consisting of MSCs and organics—effectively shielding the MSC nuclei from the reaction solution (**Figure 1**). By following the evolution of both organic and inorganic constituents, through a combined analysis of *in-situ* NMR and X-ray scattering and *ex-situ* optical spectroscopy and electron microscopy, we discovered that, upon formation of MSCs in highly concentrated solutions, long-range mesophase structures (100's nm) are formed. In contrast to previous studies at lower temperatures (<100°C),[11,12,14,21,28] we demonstrate that MSCs exist within a mesophase structure at elevated temperatures (~130°C). Our results reinforce an emerging understanding that the stability, or resistance to growth, of MSCs originates from a surfactant (or ligand) mesophase or coordination network, in spite of the atomic arrangement of cluster core. We show that high precursor concentration promotes highly selective nucleation of a single MSC species. We leverage the highly selective reaction to directly probe and track the kinetics of the MSC synthesis. Overall, high concentrations accentuate surfactant phase behavior promoting the formation of high-purity MSCs along with a stabilizing hexagonal mesophase.

**Results**

**Synthesis Concentration—**The concentration of the precursors controls the reaction pathways: at lower concentrations both NPs and MSCs are formed, while at higher concentrations MSC formation is promoted and NP growth is suppressed. We used a simplified organic synthesis involving only cadmium oleate, oleic acid, and tri-octyl phosphine sulfide (TOP=S) using a one-pot, heat-up method (see **SI for additional details**). We investigated three different cadmium oleate concentrations (100, 500, and 1000 mM), with the balance of the solution being oleic acid and TOP=S (2500 mM TOP=S; stoichiometric ratio 2:1 Cd:S). At high concentrations (500 and 1000 mM), absorption spectra of the cleaned product show a single, narrow (111 meV FWHM) excitonic peak at 324 nm (**Figure 2a**, note log scale on vertical axis). Synthesis at conventional concentrations (100 mM) does not show an excitonic peak. We previously determined the composition of these MSCs as predominately organic (70 wt%), with a 2:1 Cd:S ratio, and a repeat formula unit of $[(CdS)Cd(OA)_2]_x$ (where OA is oleic acid).[33] Based on an empirical sizing curve,[34] the peak at 324 nm corresponds to a particle size of 1.64±0.05 nm. The small deviation in size (which is approximately 1/5th of a Cd-S bond) suggests that each cluster has an identical number of Cd atoms. From published data on similar CdS clusters, we estimate the number of Cd atoms per cluster to be between 17 to 32 atoms.[35–40]

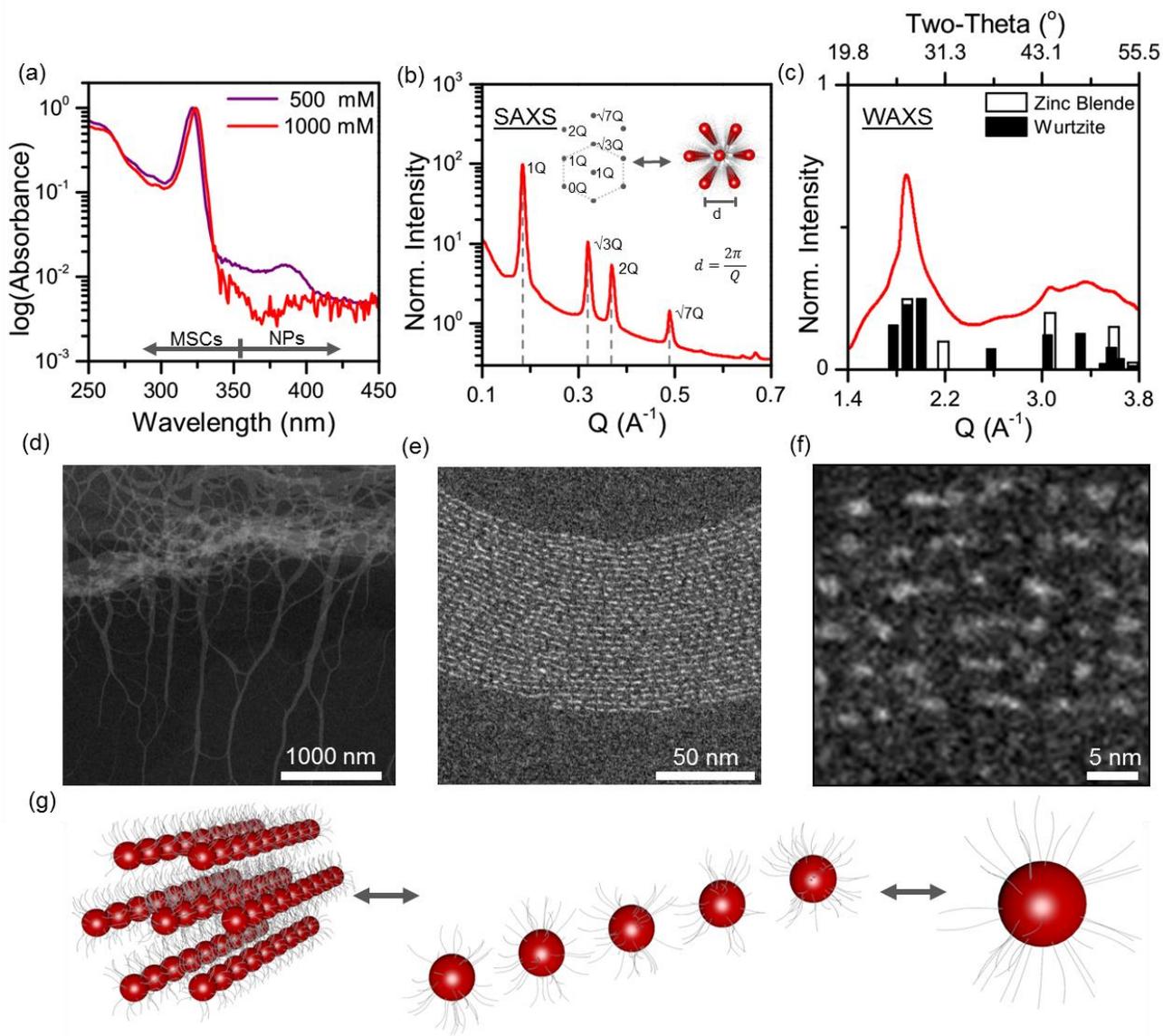

**Figure 2:** *MSC Electronic and Physical Structure.* (a) Log of absorbance for cleaned magic-sized clusters prepared at two different metal precursor concentrations (500 and 1000 mM). (b) SAXS of 1000 mM MSC synthesis at 6 h at 130°C. Inset: reciprocal and real space model of hexagonal MSC assembly. (c) WAXS of 1000 mM MSC synthesis at 6 h at 130°C compared to zinc blende (PDF#00-010-0454) and wurtzite (PDF#00-041-1049) CdS reference peaks. (d-f) STEM images of MSCs. (d) Long (> 1 μm) bundles of fibers composed of MSCs. (e) Zoomed-in view of fibers (3.4 nm d-spacing, Figure S3). (f) discrete MSCs (1-2 nm) within a fiber. (g) Schematic of the MSC hexagonal mesophase. The mesophase (left) is an assembly of nanofibers (center), which are composed of magic-sized clusters (right).

Over the course of the 1000 mM reaction (6 h at 130°C), the peak at 324 nm increases in intensity but does not shift, indicating continuous formation of MSCs (**Figure S1**). At the highest concentration, 1000 mM, the peak at 324 nm is dominant with only a small contribution from a broad NP peak (~300 meV FWHM) that shifts during growth from 375 to 404 nm (NPs account for <0.1% of total product based on particle concentration, see **SI**). For the 500 mM reaction, the NP peak that accompanies the MSC, 324 nm, peak is more intense compared to the 1000 mM reaction and located at 387 nm (FWHM ~200 meV). This broad peak is a signature of more polydisperse, continuous growth NPs; the 387 nm peak corresponds to 3.0±0.5 nm diameter NPs.[34] Over the course of the 500 mM reaction (6 h at 130°C), the MSC peak at 324 nm increases and then decreases in intensity but does not shift (**Figure S1**). This increase and decrease in intensity without a peak shift indicates the MSCs are increasing and decreasing in number but their size is not changing. Concomitantly, the reduction of the MSCs peak corresponds to a shift (358 to 418 nm) in the broad NP peak. By converting the peak intensity to concentration of MSCs and NPs, using a size-dependent extinction coefficient,[34] we can quantify the purity of the MSC product relative to the NPs (see **SI** for purity calculation; note the purity calculation assumes the empirical extinction coefficient and sizing curve are accurate for ultrasmall particles). These results show that higher precursor concentrations result in a higher purity of MSCs product, specifically 99.1% and 99.9% for the uncleaned products of the 500 and 1000 mM reactions, respectively. After cleaning (see **SI** for details) the 1000 mM reaction, there is no detection of a NP peak (purity >99.9%; see **Figure S1b**). At the same conditions, the conventional concentration (*i.e.*, 100 mM) does not produce MSCs or NPs. (Note: upon cooling from 130°C or further heating to 200°C (see **SI**) of the solution, NP precipitate.) Thus, precursor concentration tunes the synthesis selectivity from no products to high-purity MSCs.

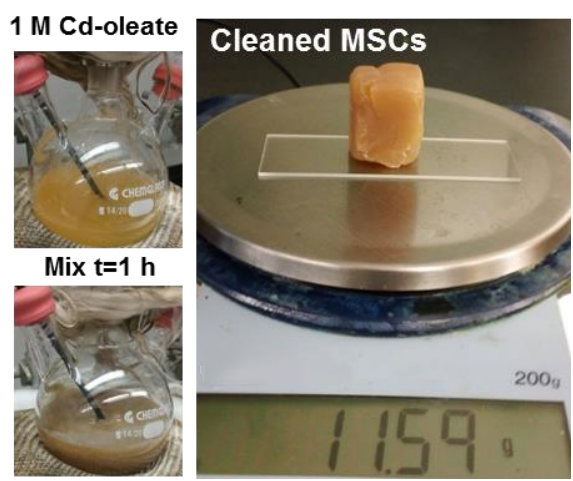

**Figure 3:** *MSC Turbidity and Scale-up.* Upon MSC formation at 140°C, the cadmium oleate + TOP=S mixture transforms from a translucent orange solution (similar to Cd-oleate only solution) to a turbid tan solution (from top to bottom). The right image is the 11.6 g of cleaned MSCs produced from a 100-mL 1000 mM reaction (~30% inorganic mass[33]). Conversion is ~25% after 1 h.

**Mesophase Structure**—The observed preference for high-purity MSCs at high concentrations raises the question: *what stabilizes MSCs in solution and what prevents the transition to NP growth?*

The answer detailed below is based on the stabilization derived from changes in the surfactant (*e.g.*, cadmium oleate) phase behavior at high surfactant concentration. Mesophase formation is suggested by the increased turbidity (caused by light scattering with large particulates) in the solution upon MSC formation (**Figure 3**). Notably, the stability of MSCs at high concentration enables their large-scale production and isolation (**Figure 3**). *In-situ* small- and wide-angle X-ray scattering (SAXS/WAXS) at 130°C for the 1000 mM reaction confirm the formation of well-ordered mesoscale assemblies (**Figure 2b**) and small crystal domains (**Figure 2c**). The combination of SAXS and WAXS is particularly powerful since SAXS captures larger mesophases while WAXS examines the individual MSCs. For the 1000 mM reaction, SAXS shows several narrow peaks while several broad peaks are detected in WAXS. The narrow SAXS peaks signatures are characteristic of hexagonal spacing in reciprocal space (Q spacing of the peaks is 1:$\sqrt{3}$:2:$\sqrt{7}$; the corresponding Miller indices are 100, 110, 200, 210, respectively; see **Figure 2b**). The first hexagonal mesophase peak (1Q ~ 0.1845 $Å^{-1}$) corresponds to a 3.4-nm d-spacing, and has an extremely large mesophase crystallite size, > 170 nm (see **SI** for details; the peak width is dominated (~86%) by instrumental broadening). The change in slope at low Q (0.1-0.2 Å$^{-1}$) corresponds to the NP structure factor. The peaks in the WAXS from the MSCs align with the diffraction planes for cadmium sulfide, and most closely with the wurtzite (WZ) phase (**Figure 2c**). For instance, the weak peak at 2.6 $Å^{-1}$ represents a characteristic (102) wurtzite plane—corresponding to 2Θ = 36.6° for a Cu-α radiation—which is absent in the zinc blende (ZB) phase. The observation of the more thermodynamically stable WZ phase under kinetically controlled conditions is consistent with computational work that show phase stability to be dependent on size and surface termination (for Cd-terminated, as is the case here, WZ is preferred),[41,42] as well as demonstrate polytypism in WZ/ZB systems depending on prepration.[43] The breadth of the WAXS peaks suggests the MSCs have a ~2 nm crystallite (**Figure S2**), which is much smaller than the mesophase grain size (>170 nm), indicating that a mesophase grain contains thousands of MSCs.

To better understand the fundamental link between the MSC stability, or their resistance to form NPs, and their mesoscale structure we imaged the cleaned 1000 mM MSC synthesis product using aberration corrected scanning transmission electron microscopy (STEM) (**Figure 2d-f**, see **SI**). Unexpectedly, the STEM images reveal long (~10's μm), fibrous assemblies with ~3 nm inter-fiber spacing (**Figure 2d-f** and **Figure S3**). The inter-fiber spacing based on STEM is consistent with the 3.4 nm d-spacing of the first hexagonal mesophase peak (**Figure S3**). Closer inspection reveals that the fibers are not continuous inorganic wires, but consist of discrete inorganic entities, each ~2 nm in size. During high resolution STEM imaging, the clusters restructure and degrade quickly under electron irradiation, which has prevented more detailed, atomic-level imaging (see **Figure S4**). Overall, the size of these entities observed via STEM is similar to the size determined from WAXS (~2.2±0.9 nm, **Figure S2**) and absorption spectroscopy (1.64±0.05 nm, based on empirical sizing curve[34]). We summarize the hierarchical arrangement of MSCs within fibers, and fibers within hexagonal mesophase in **Figure 2g**. The lack of a hexagonal mesophase in the STEM data suggests that the mesophase structure unbundles upon dilution, and indicates that the improved stabilization arises from MSCs locked into fibers rather than from the assembled mesophase.

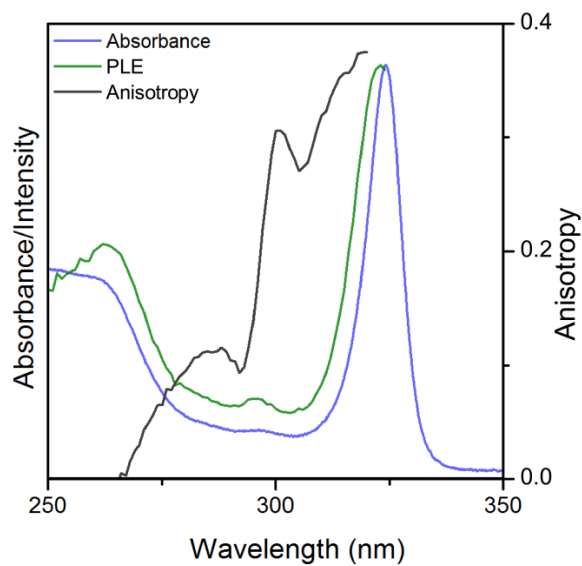

**Figure 4:** *Polarization Anisotropy.* Comparison of absorbance, photoluminescence excitation (PLE), and photoselection-PLE (PS-PLE or fluorescence anisotropy). In contrast to the isotropic emission typical in spherical NPs, MSCs show linearly polarized emission more similar to elongated nanorods. PLE was measured specifically at 331±1 nm to characterize only transitions related to the main PL peak of the MSCs.

Beyond X-ray scattering and high resolution STEM imaging, the anisotropic arrangement of MSCs within the fibrous assembly can also be probed by measuring the optical fluorescence anisotropy. Fluorescence (or polarization) anisotropy[44–47] provides a particle-level probe into anisotropy, and the response is likely not influenced by neighboring clusters given the large gap between the MSCs (~2 nm). We observe a strong anisotropic response reaching close to 0.4 near the band gap. The 0.4 value corresponds to a linearly polarized and parallel absorption and fluorescence transition dipole moment. This result indicates that the allowed electronic transitions of the MSCs are anisotropic, which implies that their crystal structure is anisotropic as well (**Figure 4** and **S5**). Comparing the MSCs anisotropy properties to larger CdS NCs with the same crystal structure shows they are more similar to elongated nanorods (NRs) with linearly polarized emission, rather than to spherical NPs that show isotropic emission in solutions. For NRs, the linear polarization arises due to their anisotropic structure,[46,48] suggesting that the shape of the MSCs is anisotropic as well. We therefore conclude that the MSCs define the basic building blocks of the fibrous hexagonal assemblies, and that the underlying anisotropy in the MSC promote the fibrous assembly.

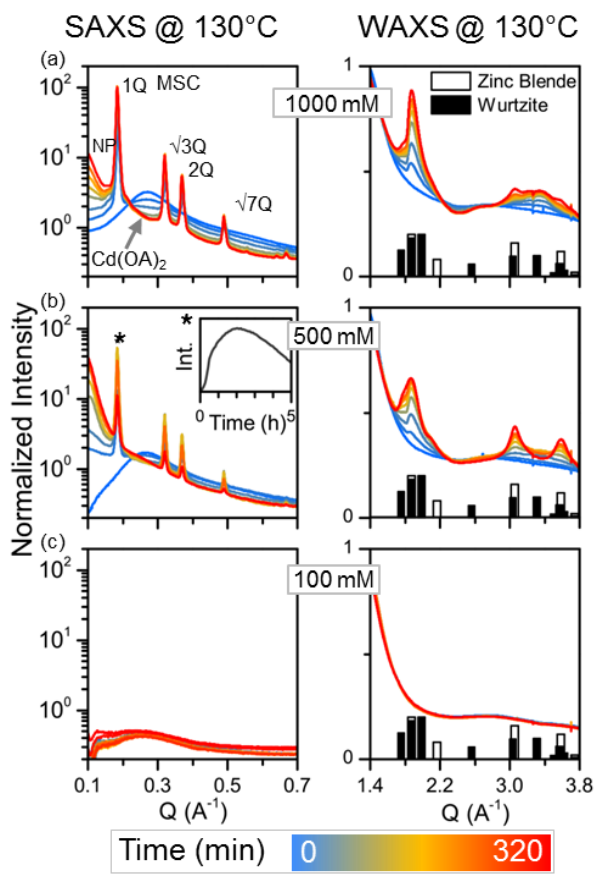

**Figure 5:** *In-situ Mesophase and MSC Formation. In-situ* SAXS and WAXS at 130°C for magic-sized clusters prepared at three different metal precursor concentrations (1000, 500, 100 mM). (a) In the 1000 mM case, the data shows formation of MSCs (WAXS) with assembly into a large hexagonal mesophase (SAXS). (b) 500 mM reaction initially forms a hexagonal assembly (SAXS), which fades at longer times when NP formation increases (WAXS). (Inset) Intensity over time for peak at 1Q, which corresponds to mesophase assembly. (c) For the 100 mM reaction no CdS reaction is observed.

*In-situ* **Mesophase and MSC Formation**—The combination of small- and wide- angle X-ray scattering (SAXS/WAXS) provides an opportunity to monitor the co-evolution of the mesophase and the constituent MSCs in real time (**Figure 5**). We investigated how the formation of the mesophase and inorganic structures is influenced by precursor concentration (100, 500, 1000 mM) over the course of 5 h at 130°C (**Figure 5** and **Figures S6-S9**). Below, we describe the co-evolution of SAXS and WAXS patterns for each concentration.

For the 1000 mM reaction, the broad SAXS peak (Q ~ 0.27 Å$^{-1}$) decays in intensity over time, while a new set of SAXS and WAXS peaks emerge and increase in intensity (**Figure 5a**). The broad decaying SAXS peak (Q ~ 0.27 Å$^{-1}$, 2-nm d-spacing) originates from the cadmium oleate precursor (**Figure S6**), which is likely a micellar phase as previously seen at high concentrations.[49] The spacings of the narrow hexagonal mesophase (SAXS) peaks stay constant while their peak intensity increases with time, indicating an increased abundance of a singular mesophase in solution. The intensity of the mesophase peaks increase concurrently with the intensity of the broad WAXS MSC peaks, suggesting that formation of MSCs is inherently coupled to the mesophase formation. Similar behavior is observed at the maximum or neat cadmium oleate concentration reaction (*i.e.*, 1580 mM at 130°C) (**Figure S10**).

By comparison, the scattering signature of the synthesis at 500 mM is similar to the 1000 mM reaction, *i.e.*, narrow SAXS and broad WAXS peaks emerge and increase with time. However, at longer times (> 2 hr) the narrow mesophase peaks begin to fade, while the low-Q (0.1-0.2 Å$^{-1}$) structure factor changes slope, and sharper WAXS peaks appear (**Figure 5b**). The increasing slope at low-Q indicates an increase in the particle size (**Figure S11**). Interestingly, the sharper WAXS peaks align better with the ZB phase of CdS, in contrast to the WZ-like structures observed for the MSCs at high concentrations. Specifically, the ZB peaks at 3.1 and 3.6 Å$^{-1}$ ((220) and (311) diffraction planes; 44.0° and 52.1° 2Θ for Cu K-α source, respectively) appear while a WZ peak at 3.3 Å$^{-1}$ ((103) plane; 47.8° two-theta for Cu K-α source) fades.

At conventional concentrations (100 mM), no change in intensity or peak formation is observed in the SAXS or WAXS (**Figure 5c**). The transformation between the fibrous mesophase and NP growth at 500 mM implies the higher selectivity for MSC at 1000 mM is a result of the MSC fibrous assemblies, which likely impede the onset of the NP growth.

The evolution of the MSC and mesophase structure evident from the SAXS/WAXS results summarized in **Figure 5**, led us to hypothesize that the stability of the MSCs in the high-concentration environment derives from the clusters being assembled into fibers and encapsulated within fibrous surfactant mesophase. The arrangement of MSCs within the fibrous mesophase limits the mobility of the nuclei and effectively freezes or isolates them from the surrounding reaction environment. Whereas the conventional mantra has been that the stability of MSCs derives from the precise arrangement of the inorganic structures (*i.e.*, closed-shell atomic packing),[23,24,50] we demonstrate that anisotropic clusters can also be stabilized within fibrous assemblies. We note that the lack of strong order within the MSC building blocks studied here is consistent with a recently isolated stable nanocluster that has a disordered structure, and is hypothesized to be stabilized by interconnected networks of surface ligands.[29,51] Collectively, these results underscore that higher concentrations promote not only the formation of MSCs in high purity, but also generate liquid-crystalline fibrous mesophase assemblies. These assemblies afford an additional level of stabilization for the clusters at high concentrations.

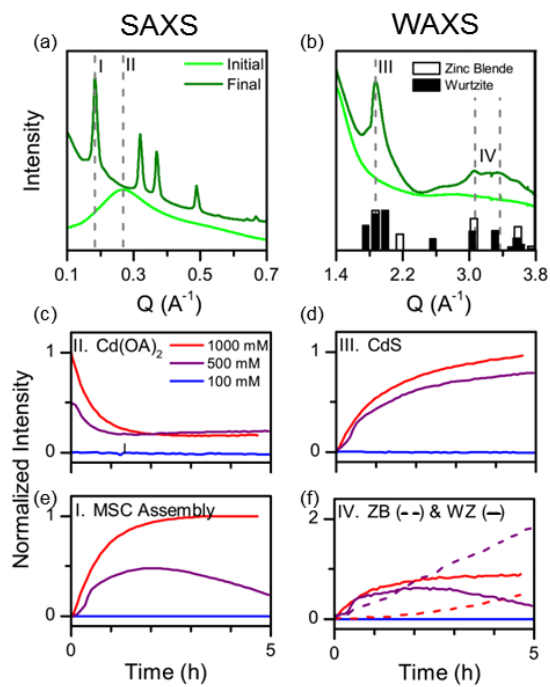

**Figure 6:** *Time-resolved X-ray Analysis.* Comparison of initial and final patterns for 1000 mM synthesis at 130°C, (a) SAXS and (b) WAXS with enumerated characteristic peaks (I-IV). (c-f) Comparison of the evolution of reactant and product peaks at 130°C for three different precursor concentrations: 100, 500, 1000 mM (see SI for details on analysis). (c) The Cd oleate micelle peak (II) decreases, and then levels off over the course of the reaction. (d) CdS diffraction peak (a shared ZB and WZ peak, 1.87 Å$^{-1}$, labeled (III)), increases with time indicating that the loss of Cd-oleate corresponds to the formation of CdS. (e) The intensity of the sharp 1Q hexagonal MSC assembly peak (I) increases with time for the 500 mM and 1000 mM reactions, but not for the 100 mM reaction. This peak begins to fade away at longer times in the 500 mM, but not the 1000 mM reaction. (f) Intensity of characteristic ZB (- -) and WZ (—) peaks at 3.07 and 3.37 Å$^{-1}$, respectively (labelled (IV)). For the 1000 mM, the formation of the MSCs mirrors the WZ peak intensity. For 500 mM, the loss of WZ MSC intensity with time mirrors the decay of the MSC assembly, and is accompanied by linearly increasing ZB intensity from NPs growing at the expense of MSCs. While the ZB peak overlaps with a WZ peak, the fact that this peak increases, while a unique WZ peak decreases, suggests the increase in peak intensity is due to ZB formation.

By tracking characteristic peaks in SAXS and WAXS during the synthesis we can directly compare the structural evolution of the inorganic discrete MSCs and their assembly into ordered fibrous ensembles during synthesis (**Figure 6**). To disentangle the complex interplay between simultaneous atomic, nanoscale, and mesoscale phenomena, we focus on four characteristic peaks corresponding to: (I) the first peak for the hexagonal assembly, (II) the cadmium oleate micelles, (III) CdS diffraction peak shared by both ZB and WZ phases (1.87 Å$^{-1}$), and (IV) characteristic ZB and WZ peaks (at 3.07 and 3.37 Å$^{-1}$, respectively) (**Fig. 6a,b** and **SI**). To highlight the critical effect of the precursor concentration, we compared the 100, 500, and 1000 mM reactions. The 100 mM reaction shows no peak change or MSC formation, hence the normalized intensity is zero and no further analysis was performed. The cadmium oleate micelle peak (II) is initially more pronounced in the 1000 mM reaction, and decays exponentially in both the 500 mM and 1000 mM reaction (**Figure 6c**). This indicates that more cadmium oleate micelles form in the 1000 mM as compared to the 500 mM reaction, which is not surprising given the higher concentrations. Notably the cadmium oleate peak for 1000 and 500 mM level off at roughly the same time. Concurrently, the hexagonal MSC mesophase peak (I) increases (**Figure 6e**) along with the slope of the NP structure factor peak at low-Q (**Figure S11**). In the 1000 mM reaction the hexagonal MSC peak (**Fig. 6e**, red plot) rises and ultimately plateaus in time; but, surprisingly, for the 500 mM reaction this peak reaches a maximum, and then decays with time (**Fig. 6e**, purple data set). The mesophase structure is roughly twice as abundant in the 1000 mM reaction compared to 500 mM at 2 h.

These real-time *in-situ* studies reveal that the formation of the MSC SAXS assembly peak coincides with the formation of several broad peaks in the WAXS, which are characteristic CdS diffraction planes. The loss of MSC assemblies in the 500 mM reaction, but not for the 1000 mM, highlights the stability supplied by the hexagonal surfactant mesophase. This stability degrades upon prolonged exposure to higher free acid concentrations (in the 500-mM reaction), akin to an Ostwald ripening mechanism for NP, or exposure to higher temperatures, which promote NP growth (see **Figure S12** 1000 mM reaction at 175°C). While previous results suggest the closed-shell stability of MSC prevents Ostwald ripening (partial dissolution),[30,52] the loss of MSCs at 500 mM implies that individual MSCs completely dissolve into monomers and surfactants to grow NPs. While both conditions (500 and 1000 mM) form CdS MSCs with mesophase assemblies, the 1000 mM concentration is significantly more resistant to mesophase degradation and NP growth.

Closer inspection of the WAXS measurements provides important insight into the transformation from MSC nucleation to NP growth. We compared the formation and phase of the CdS by tracking a shared ZB/WZ peak and characteristic ZB-only and WZ-only peaks. **Figure 6d** shows that over the same reaction time, broad CdS diffraction peaks (III) increase for both the 500 mM and 1000 mM reactions, but more significantly for the 1000 mM reaction. The growth of the WAXS signature (**Figure 6d**), indicating the formation of small CdS crystallites, and the concomitant decay of the Cd oleate micelle signature (**Figure 6c**) suggest that the two processes are related. As previously discussed, the MSCs are more WZ-like whereas the NPs are more ZB-like. At longer times (>1 h), the intensity of the ZB peak increases due to NPs growth (**Figure 6f,** dashed lines). The ZB peak intensity increases more rapidly in the 500 mM compared to the 1000 mM reaction (**Fig. 6f**, dashed curves). Over the same time (>1 h) in the 500 mM reaction, the characteristic WZ peak of the MSCs begins to decay, whereas the peaks plateau in the 1000 mM reaction. (**Fig. 6f**, solid curves). Hence, ZB-like NPs form not only at the expense of the MSC assemblies (**Figure 6e**), but also at the expense of the MSC WZ-like inorganic phase (**Figure 6f**). These results highlight that hexagonal MSC assemblies provide a barrier to NP growth that is more pronounced at higher concentration.

Discussion

Access to high-purity MSCs enables us to directly address two fundamental questions with regards to control of MSCs: (1) What is the source of MSC stability against growth? and (2) What are the factors governing the formation of either MSCs or NPs?

**MSC Stability and Mesophase Formation**—The time-resolved SAXS/WAXS data described above provide new insights into the origin of MSCs stability. Conventionally, MSCs have been suspected to be stabilized by a symmetric, close-shelled structure that resists atom-by-atom addition (*i.e.*, no low coordination atoms).[23,24,50] However, several groups have demonstrated that different MSCs correspond to differences in surface chemistry rather than size.[33,53] Moreover, other studies have shown that same-sized MSCs can be structural polymorphs[54] or possess atomically disordered structures.[51] While the small size and intermediate nature of a MSC make it difficult to resolve the underlying structure, these results suggest that the structure of the inorganic core alone may not be the origin of MSCs stability.

In this work, we demonstrate that there is a strong link between a fibrous assembly and MSC stability, by showing that mesophase assembly accompanies MSC formation. Subsequent resuspension and heating of cleaned MSCs highlights that the individual fibers rather than their macroscopic mesophase assembly are the source of MSC stability (see discussed below). The link between MSC self-assembly and stability is reinforced by the observation that degradation of the fibrous mesophase assembly destabilizes the MSCs, and results in the loss of MSCs and the enhancement of NP growth (**Figure 6**). At high concentration, the fibrous MSCs assembly is retained, and shields (*i.e.,* kinetically arresting) the MSCs against growth. Furthermore, the MSCs are also locked in at a single size, and stabilized against the quantized growth that is often observed between different MSC families (**Figure 7**).[22,27,30,55] We hypothesize that the MSC fibrous assemblies may be the source of MSCs stability in addition to a symmetric inorganic structure as has been previously proposed.

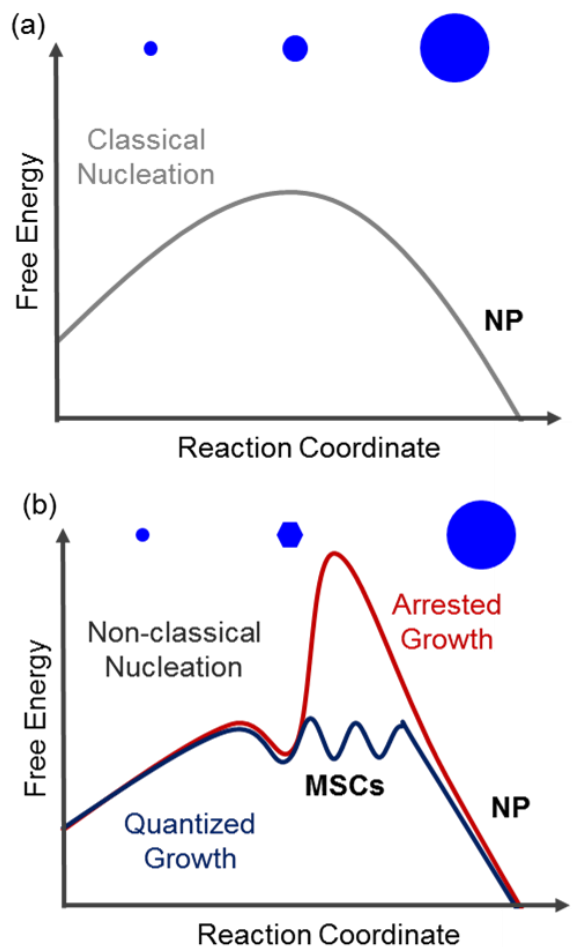

**Figure 7:** *MSC vs. NP Reaction Coordinate.* Qualitative comparison of (a) classical and (b) non-classical nucleation barriers along with growth barriers for quantized MSC growth and the kinetically arrested growth. The formation of a mesophase generates an additional energy barrier, restricting the transition from MSCs formation to NP growth, and locking the MSCs in the nucleation phase.

To investigate the mechanism of growth from (or destabilization of) MSCs into NPs without precursors, (*i.e.,* through Ostwald ripening and/or coalescence),[27] we heated the MSCs in two forms— (1) cleaned and resuspended in 1-octadecene (ODE), and (2) a cleaned solid product (no resuspension in solvent). We monitored the system with SAXS/WAXS. At 100°C, the concentrated solution of MSCs in ODE (100 mg/mL) did not show a hexagonal mesophase, suggesting that the mesophase completely unbundles upon solvent intercalation to form a fibrous suspension (**Figure S13**). The persistence of a fibrous suspension is suggested by highly viscous and gel-like solution behavior, and reinforced by the presence of large structural features observed using dynamic light scattering (**Figure S14**). Heating the resuspended MSC solution to 200°C did not induce NP growth, highlighting the remarkable stability of the MSCs against growth either by coalescence or ripening. (**Figure S13**). The resistance of cleaned and resuspended MSCs to growth and dissolution is consistent with a previously published study.[30] MSC stability, in the absence of a mesophase, indicates that the individual fibers of MSCs, and not their mesophase assemblies, are the fundamental source of MSC stability.

For the cleaned solid MSC product experiment (*i.e.*, without solvent), the hexagonal mesophase is observed and retained upon heating to 200°C, though the mesophase assembly expands by 3% (see **Figure S15**). Both results of these heating experiments highlight the thermal stability of the fibrous MSCs. The solid-like (high viscosity) nature of hexagonal mesophase, along with the gel-like nature of MSCs in ODE, deters MSC dissolution and growth. Based on these monomer-free experiments, we conclude that at typical synthesis temperatures (*i.e.*, 100 – 200°C), MSC to NP conversion is not through cluster assembly or a coalescence mechanism when a mesophase or fibrous assembly is present.

**Surfactant Phase Behavior in Prior MSC Literature**—Several other studies indirectly allude to a connection between surfactant structure and MSC formation. Mesophases have been identified by turbid solutions,[15] and well-defined peaks in the small-angle X-ray scattering,[16,17] and/or optical birefringence.[18] In line with these metrics, several works mention the following in connection with MSC formation: solution turbidity and MSCs self-assembly,[19] and X-ray or NMR detection of large (~1 nm) micellar aggregates.[20,21] Nevertheless, these works do not connect solution or mesophase structure to the stability of the MSCs.

There may be a more fundamental connection between MSCs and mesophases through supersaturation. Previous studies indicate that high levels of monomer supersaturation promote MSC formation.[22,26] Based on the micelle theory, high levels of surfactant supersaturation (relative to critical micelle temperature/concentration) promote micelle or mesophase formation. Thus, sufficiently low temperatures or high concentrations may promote both the formation of MSCs and mesophases. Previously, the high supersaturation needed for MSC formation is achieved by lower synthesis temperatures (in some cases, <100°C),[11,12,14,21,22,28,29] and low acid-to-metal ratio (~3).[19,25,30–32] For example, at a low synthesis temperature (<100°C), Buhro and co-workers explic-itly mention that CdSe MSCs form within a lamellar surfactant mesophase composed of the precursors.[11,12,14] Another recent low-temperature study demonstrated that perovskite nanoclusters (CsPbBr$_3$) are characterized by a milky solution, and stabilized by mesophase formation.[56] Together with our results, these findings generalize the importance of mesophase formation and stabilization across a diverse set of cluster syntheses.

At low temperature or low acid concentrations, interactions between metal surfactants are intensified, and may promote the formation of mesophase assemblies. Additionally, as suggested by micelle theory, high concentration can also promote mesophase formation. A recent paper showed that using high concentration, 'solvent-free' conditions directs the reaction pathway from NP to nanoplatelet formation, and no mesophase was observed.[57] Three other studies synthesized metal carboxylates, under what we classify as high concentrations (500, 570, and 830 mM), but then diluted the precursors to conventional concentrations prior to MSC synthesis (to 120, ~120, and 250 mM, respectively), and did not mention the high concentration preparation as necessary for MSC formation—even though surfactant behavior is concentration-dependent with high concentrations promoting micelle and mesophase formation.[21,30,31] Taken together these results imply that many previous MSC studies may have been, unknowingly, observing mesophase or surfactant structure formation coupled with MSC formation.

**MSC *vs.* NP Formation**—The ratio of organic surfactant to metal plays a critical role in switching the synthetic pathway between MSCs and NPs. Based on our results (**Figure 6** and **S16**) and previous literature reports, the estimated crossover point from NP to MSC formation is an acid/Cd ratio of <16, which corresponds to a free acid vs. total acid or oleate percentage of <87%.[32,58] Specifically, we find that for acid-per-metal ≤16 (*i.e.*, concentration ≥200 mM), MSCs form in larger number than NPs. Previous reports mention that low acid-to-metal ratio (~3), even at low precursor concentrations (20 mM, diluted with ODE),[19,30–32,58] promote MSC formation, and MSCs are detected up to an acid-to-metal ratio of 10.[32,58] At lower acid/Cd ratios, there is less free surfactant to stabilize/disperse the metal precursor; thus, precursor-precursor (*i.e.*, cadmium oleate) self-interactions are preferred over solvent/surfactant interactions, promoting the formation of precursor solution structure (*e.g.*, micelles). We propose that these precursor-precursor interactions provide a stronger driving force for MSC formation over NPs, and are more prevalent at high concentrations. While low acid/Cd ratios, even at dilute conditions, are sufficient to promote MSC formation, only high concentrations suppress NP growth (**Figure S17**).

We observe that MSCs transition into NP at longer times, high temperatures, or upon addition of coordinating solvents to the 1000 mM reaction. Several other common synthesis parameters (stoichiometry, ramp rate, and ligand length) did not affect the 324-nm MSC formation, hexagonal mesophase formation, or MSC into NP growth (see **Figures S18-S23**). As was previously shown (**Figure 5**), both 500 and 1000 mM cadmium oleate concentrations generate MSCs and a hexagonal mesophase. At longer reac tion times for the 500 mM reaction, the mesophase and MSCs decay while NPs grow (see **Figure 5** and **Figure S1**), resembling the Ostwald ripening mechanism which is known to occur at high free acid concentrations.[4] Increasing the reaction temperature also leads to the formation of NPs, and the loss of the hexagonal mesophase and MSCs, but no transformation to a different mesophase structure (**Figure S13**). These results indicate that MSCs destabilize, and degrade, as they transition to NPs, and that MSC to NP conversion is predominately through monomer-driven growth, not cluster coalescence.

**Precursor or MSC Templated Mesophase Formation** — The driving force behind the mesophase formation is the assembly of MSCs and/or precursor templating. The anisotropic shape of MSCs (**Figure 4**) and the fact that independent of precursor chain length, identical fibrous MSCs form (**Figure S22**) suggests that the inorganic core is driving the fibrous assembly. Nevertheless, there is some contribution from the cadmium oleate. To better understand the role of free solvent on cadmium oleate structure, we cleaned 1000 mM cadmium oleate from the free oleic acid, and resuspended the neat cadmium oleate at 1000 mM concentration in several different solvents: ODE, oleylamine, trioctylphosphine oxide (TOP=O), and dodecanol. Then, TOP=S was injected and the solution was heated to 130°C. Reaction in a non-coordinating solvent (ODE) yields similar results to the original 1000 mM synthesis reaction: only MSCs are formed along with a hexagonal assembly (**Figure S24-S26**). For coordinating solvents, only the dodecanol formed MSCs with a hexagonal assembly, and also some NP growth, whereas oleylamine and TOP=O produce mainly large NPs without any significant mesophase (**Figure S24-S26**). The addition of coordinating solvents disrupts the cadmium oleate coordination,[59] and reveals that the solution structure of not only the MSCs but also the cadmium oleate precursor is essential to stabilize the MSCs, and the mesophase, and deter NP growth. Recent studies have shown that coordinating solvents also disrupt the ligand networks of individual MSCs,[33,60] indicating that the surfactant/ligand structure is intimately connected with both the formation, stability, and assembly of MSCs.

Metal carboxylates such as cadmium oleate precursor are well-known within other fields to exhibit solution structure and are classified as heavy metal soaps/surfactants,[6,9,61,62] metallomesogens,[17,49,63–65] metallogels,[66–69] and/or coordination polymers.[59,66,69] Cadmium carboxylates typically form solution structures that can be described as columnar or worm-like micelles, coordination polymers, fibrous metallogel and/or hexagonal mesophases.[10,59,61,69,70] We observe evidence of micelle structure cadmium oleate at 130°C with a 2.6-nm d-spacing (Q ≈ 0.25 Å$^{-1}$) (**Figure S7**). In contrast to the small micellar size of the cadmium oleate (Sherrer size ~ 10 nm; **Figure S7**), the MSC mesophase emerges as an extremely narrow peak (100's nm grain) rather than narrowing as the reaction proceeds. This behavior reinforces the idea that cadmium oleate exists as long, worm-like micelles or as a coordination polymer prior to MSC formation. The addition of coordinating solvents alters the micelle structure, and directs the synthesis away from MSCs, and toward NP formation (**Figure S26**). A recent *in-situ* SAXS study (dilute 30 mM cadmium myristate + chalcogenide source at 100°C) did not show a peak around Q=0.25 Å$^{-1}$, and reports < 1 nm sized micelle.[71] Therefore, only higher concentrations of cadmium oleate increase the interaction probability between individual cadmium oleate surfactant creating larger micelles (or coordination polymers) that template MSC formation and assembly. If these precursor structures are disrupted (with coordinating solvents), the MSCs are destabilized, and NP growth ensues. Based on these results, we conclude that both the anisotropic shape and precursor structure template the mesophase formation.

**Kinetics**—Beyond insights into the MSC stability and MSC *vs*. NP selectivity, the time-resolved SAXS/WAXS experiments also provide new understanding into the kinetics of MSC and NP formation. It is generally believed that colloidal syntheses require burst nucleation to produce monodisperse particles, and nuclei are treated as unstable and fleeting transition states, quickly overtaken by NP growth.[2,3,72] In contrast, non-classical syntheses may involve continuous nucleation, and stable and persistent intermediates, or clusters, can be observed.[5,27,29] The ability to isolate and track these crucial reaction intermediates, which often can be MSCs, provides a probe into the non-classical NP synthetic pathway (and nucleation processes) in a way not achievable in conventional NP syntheses. Specifically, the selectivity for MSCs over NPs, at high concentrations, decouples precursor conversion kinetics that contribute to MSC formation from NP growth.

We probed the precursor kinetics for MSCs using *in-situ* NMR and X-ray scattering and *ex-situ* absorbance analyses. By following the organic precursor constituents via NMR we observe a similar precursor conversion mechanism as reported previously by Owen and co-workers,[73,74] in which metal carboxylates react with trioctylphosphine chalcogenides to form TOP=O, oleyl anhydride, and metal chalcogenide monomer (**Figure 8** and **S28-S29**). The resulting metal chalcogenide monomer then nucleates and grows to form MSCs and/or NPs. However, a notable difference in our high concentration synthesis is that some species become NMR silent (*i.e.,* the total spectral area decreases) over the course of the reaction, based on $^{13}$C, $^{31}$P and $^{1}$H NMR (**Figure 8a**). This NMR silencing indicates solidification,[75–77] and this process occurs at a similar rate as the formation of the MSCs mesophase (~$10^{-4}$ s$^{-1}$, see **Table S5** and **Figure S30**). While the precursor conversion products (*i.e.,* TOP=O and oleyl anhydride) match those in conventional synthesis, synthesis at high concentration involves another process (*i.e.,* phase change, observed as NMR "silencing") that promotes highly selectivity MSCs formation and assembly.

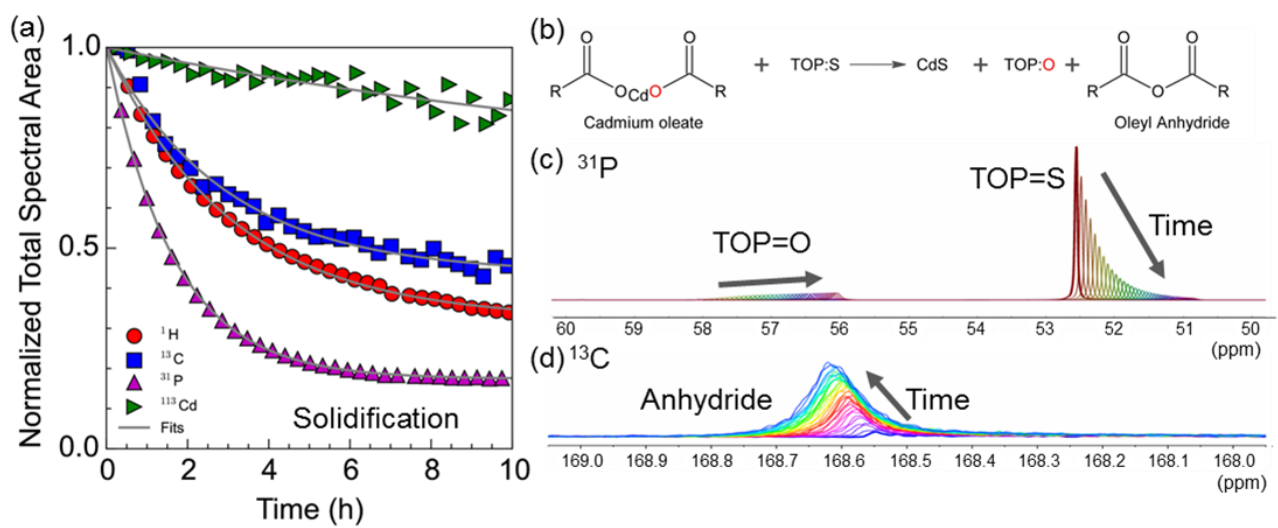

**Figure 8:** *1000 mM Synthesis at 130°C (NMR).* **(a)** The total spectral area of the four nuclei investigated decreases with time (H, C, P, and Cd). The silencing of the signal is attributed to species becoming NMR inactive through solidification.[75–77] (see Table S5). **(b)** Reaction schematic showing the conversion from precursor (Cd-oleate and TOP=S) to CdS along with organic by-products (TOP=O and Oleyl Anhydride). **(c)** $^{31}$P NMR spectrum showing the emerging peak (between 58 and 56 ppm), which is assigned to TOP=O formation while the initially more intense, but decreasing, peak (between (52.5 and 50.5 ppm) is assigned to TOP=S (from literature 48.5 to 50.2 ppm values are observed for TOP=O and 48.6 pm for TOP=S, respectively).[73] It is known that P peaks can shift in the presence of Lewis acid (which could include oleic acid and possible cadmium oleate); **(d)** $^{13}$C NMR spectrum showing the emergence of an oleyl anhydride peak (($-CH_2CO)_2O$ ) that increases with time, and is at a similar location (168 ppm) to previously reported oleyl anhydride peak.[73]

Regarding the inorganic constituents (MSCs and NPs), the *ex-situ* absorbance spectra show a linear or zero-order reaction for the MSC formation (2±0.1 x $10^{-6}$ M s$^{-1}$; see **Figure S30**). This rate is slower, but similar to precursor conversion during the initial nucleation phase of CdSe NPs at higher temperatures ($10^{-5}$ M s$^{-1}$),[4] and much slower than the first-order precursor conversion rate that includes NP growth ($10^{-3}$ -$10^{-1}$ s$^{-1}$).[1,72,73] The linear (or zero-order) relationship between precursor conversion rate and NP production is expected based on classical nucleation theory during the nucleation phase,[1,6,78] indicating that the MSCs are similar to nuclei.[27,29] Nevertheless, our MSCs are locked at a single size and do not grow in contrast to NP nuclei that continuously grow. We observe continuous nucleation of MSCs over an extremely long time (~6 h) compared to typical burst nucleation times (seconds) in conventional synthesis. The prevalent classical understanding is that the burst (short-lived nucleation) is crucial to obtain monodisperse NPs;[3,78] In stark contrast, we show that high concentration synthesis promotes continuous nucleation of clusters and deters NP growth via fibrous assembly and mesophase formation, ultimately supplying a batch of monodisperse clusters.

At longer reaction times, the reaction transitions from MSC nucleation to NP growth for 500 and 1000 mM. The rate of NP formation from *ex-situ* absorbance is first order for the 500 mM reaction (6 x $10^{-4}$ $s^{-1}$) whereas no NP formation rate is observed for the 1000 mM reaction (**Figure S30**). These rates are slower than those previously reported (0.001-0.1 $s^{-1}$),[1,72,73] though literature values include both nucleation and growth contributions to precursor conversion, and are in less viscous synthesis environments. Generally, the slower rates at high concentrations align with the current understanding of the precursor to monomer conversion leading up to nucleation as the rate limiting reaction step (nucleation rates are slower than growth rates[1,2]) and the lower solution diffusivity at high concentration limits the growth rate.[79–81] The formation of mesophase structure effectively minimizes the solution diffusivity, and thus NP growth, through the formation of a solid (low diffusivity) phase. In effect, the fibrous and mesophase structures at high concentration kinetically arrest or freeze the MSCs, further stabilizing their kinetically persistent structure.

Conclusions

In summary, we demonstrate that colloidal NP synthesis in the high concentration regime accentuates surfactant phase behavior leading to the formation of high-purity MSCs stabilized within a highly ordered hexagonal mesophases assembly. Our results indicate that the fibrous MSC assemblies are likely templated from structure inherent to the cadmium oleate precursor as well as inherent shape anisotropy of the MSCs. We present a mechanism in which MSCs in the assembly are shielded from further growth (i.e., kinetically arrests NP growth), and propose that MSC assemblies may be the source of MSCs stability rather than, or in addition to, a precise inorganic structure. NP growth can be initiated at the expense of both the MSC and their hexagonal mesophase. NP growth at expense of MSC implies that synthetically MSCs are intermediates or "monomer reservoirs" for NPs rather than NP nuclei. Whereas syntheses at conventional concentrations are governed by monomer-addition-based growth, or monomeric surfactant/precursor environments, we establish that high precursor concentrations expand the colloidal phase diagram for NP synthesis into more complex surfactant phase behavior, namely micelles and mesophases. While inorganic phase change (e.g., nucleation) is fundamental to NP synthesis, the importance of innate organic phase change (e.g., surfactant mesophase formation) has been latent and underappreciated. This organic phase behavior is revealed in the new high concentration regime that is characterized by maximizing precursor-precursor interactions to form a solution structure that selectively navigates the synthetic pathway to isolate high-purity MSCs. In contrast, conventional concentrations are more sensitive to other chemical interactions with the precursors (including solvent, surrounding defects, impurities). High concentration NP syntheses selectively control the predominant molecular interactions during NP syntheses. Insight into inherent surfactant phase behavior of NP precursors as metal soaps provides a generalized framework for metal chalcogenide and perovskite NP synthesis. Overall, the high-concentration synthesis regime accentuates fundamental surfactant phase behavior, and offers a generalized method for synthesizing, stabilizing, and studying high-purity metal chalcogen clusters than are persistent intermediates in non-classical NPs syntheses.

## ASSOCIATED CONTENT

**Supporting Information**. Synthesis procedure and characterization methods. X-ray analysis methods (including peak tracking, structure factor, Scherrer analysis). Absorbance and SAXS/WAXS for 500 and 1000 mM time series. Additional STEM images. 1000 mM synthesis at higher temperature and with different free solvents (SAXS/WAXS). SAXS/WAXS while heating resuspended MSCs and solid MSCs. High concentration synthesis with different carboxylic acid ligands (SAXS/WAXS). Kinetic fits to absorption, x-ray and NMR data. This material is available free of charge via the Internet at http://pubs.acs.org.

## AUTHOR INFORMATION

Corresponding Author

*Tobias Hanrath (E-mail: th358@cornell.edu)
*Richard D. Robinson (E-mail: rdr82@cornell.edu)

Author Contributions

All authors contributed to the interpretation of results and preparation of the manuscript. All authors have given approval to the final version of the manuscript.

§These authors contributed equally.

Notes

The authors declare no competing financial interest.

## ACKNOWLEDGMENT

This work was supported in part by the National Science Foundation (NSF) under Award No. CMMI-1344562. U.B. acknowledges funding for this project from the European Research Council (ERC) under the European Union's Horizon 2020 research and innovation programme (grant agreement n° 741767). U.B. also thanks the Alfred & Erica Larisch memorial chair. B.H.S. and L.F.K. acknowledge support by the Packard Foundation. B.H.S was supported by NSF GRFP grant DGE-1144153. This work also made use of the Cornell Center for Materials Research Shared Facilities, which are supported through the NSF MRSEC (Materials Research Science and Engineering Centers) program (Grant DMR-1719875). The FEI Titan Themis 300 was acquired through NSF-MRI-1429155, with additional support from Cornell University, the Weill Institute and the Kavli Institute at Cornell. This work includes research conducted at the Cornell High Energy Synchrotron Source (CHESS), which is supported by the National Science Foundation and the National Institutes of Health/National Institute of General Medical Sciences under NSF award DMR-1332208. Dynamic light scattering measurements were performed in a facility supported by Award No. KUS-C1-018-02, made by King Abdullah University of Science and Technology (KAUST). This work made use of the Cornell Chemistry NMR Facility, which is supported in part by the NSF-MRI grant CHE-1531632.

The authors would like to thank the following individual for assistance with experiments and material characterization as well as useful discussion. Specifically, Detlef Smilgies provided equipment and helpful discussion regarding the X-ray scattering experiments. Stan Stoupin set-up and calibrated the beamline and detector for SAXS/WAXS, and helped with data analysis. Ivan Keresztes performed the NMR data acquisition, helped with analysis, and provided useful guidance.

## ABBREVIATIONS

NP, nanoparticle; MSC, magic-sized cluster; ODE, 1-octadecene; TOP=O, tri-octyl-phosphine oxide; TOP=S, tri-octyl-phosphine sulfide; SAXS, small-angle x-ray scattering; WAXS, wide-angle x-ray scattering; STEM, scanning transmission electron microscopy; ZB, zinc blende; WZ, wurtzite.

**Magic-sized Clusters**   **Mesophase**

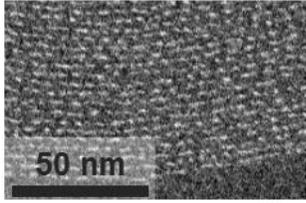
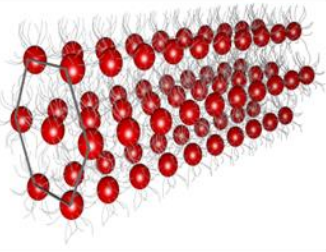
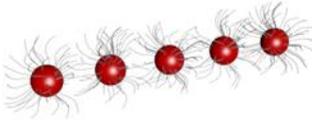

# Mesophase Formation Stabilizes High-Purity Magic-Sized Clusters

*Douglas R. Nevers[†,§], Curtis B. Williamson,[†,§], Benjamin H. Savitzky[+], Ido Hadar[#], Uri Banin[#], Lena F. Kourkoutis[‖,⊥], Tobias Hanrath*[†], and Richard D. Robinson*[‡]*

[†]Robert F. Smith School of Chemical and Biomolecular Engineering, Cornell University, Ithaca, New York, 14853, United States

[+]Department of Physics, Cornell University, Ithaca, New York, 14853, United States

[#]The Institute of Chemistry and Center for Nanoscience and Nanotechnology, The Hebrew University of Jerusalem, Jerusalem 91904, Israel

[‖]School of Applied and Engineering Physics, Cornell University, Ithaca, USA. [⊥]Kavli Institute for Nanoscale Science, Cornell University, Ithaca, New York, 14853, United States

[‡]Department of Materials Science and Engineering, Cornell University, Ithaca, New York, 14853, United States

# Table of Contents

# Synthesis Procedures

**Materials**

The following chemicals were used as received and were purchased from Sigma-Aldrich: 1-octadecene (ODE, >90%), oleic acid (OA, >90%), cadmium oxide (>99.5%), ethyl acetate (99.5%) and elemental sulfur (purified by sublimation, particle size~100 mesh), tri-n-octylphsosphine (TOP, 90%), tri-n-octylphosphine oxide (TOP=O, 99%), oleylamine (OLA, 98%), dodecanol (DDA, 98%), hexanoic acid (99%), lauric acid (98%), stearic acid (95%), and acetonitrile (ACN, 99.8%). Hexanes and Toluene (BDH ACS Grade) and ethanol (Ethanol, 200 proof, Anhydrous KOPTEC USP) were purchased from VMR International. Octanoic acid (98%) was purchased from SAFC, and erucic acid (>85%) was purchased from TCI America.

**Synthesis of ultra-pure 324-nm cadmium sulfide MSCs (1000 mM)**

For the 1000 mM reaction, we used the same preparation procedure previously reported to synthesis 324-nm magic-sized clusters (MSCs),[1] which are ultra-pure due to the lack of nanoparticle (NP) growth. For subsequent/control experiments, reaction times and temperatures were varied as described in the text. Briefly, 1000 mM cadmium oleate (Cd oleate) was prepared by heating cadmium oxide (1.28 g, 10.0 mmol) and oleic acid (10 mL) in a round-bottom flask under nitrogen. The solution was held at 160 °C under nitrogen for 1 h until clear with translucent tan color. Then, the solution was cooled to 100 °C and placed under vacuum until bubbling stopped (~ 1 h). The 2500 mM TOP=S was prepared by dissolved elemental sulfur (0.8 g) into TOP (10 mL) in nitrogen glove box. The solution dissolved quickly (~1 min), and generates significant heat. (*NOTE: Caution should be taken in scaling-up this reaction.*) Next, the Cd oleate solution was cooled to 50 °C, and the 2500 mM TOP=S was injected (2 mL; 2:1 Cd:S ratio). The Cd oleate + TOP=S mixture was heated typically heated to 130 °C. Upon formation of MSCs, the solution becomes turbid. The solution was quenched after a specified time (typically 1 h, but can be extended to at least 6 h with yield improving continuously with time) with ethyl acetate (1:1 by volume), and centrifuged at 4400 rpm for 5 min. Only small amounts of anti-solvent are required to precipitate the MSCs because of the higher relative density of MSC assemblies compared to discrete NPs. Upon resuspending in hexane, the solution gels significantly. Once dissolved, the solution was precipitated again with ethyl acetate (1:1 by volume) and centrifuged. After cleaning, the sample was dried under vacuum, producing a waxy, resin-like solid product. Notably, the 324-nm MSCs slowly convert into a 313-nm family upon expose to air, and as such should be stored in inert atmosphere or in solution at elevated temperatures (60 °C). Understanding this conversion to a 313-nm MSC is part of a forthcoming publication.

For the *ex situ* absorption experiments (500 and 1000 mM), the temperature was controlled with a hot oil bath rather than heating mantle (to mimic the near isothermal temperature profile of the *in situ* NMR and X-ray scattering experiments). The set-point temperature (130 °C) was reach within 5-10 min of inserting the flask.

**Synthesis cadmium sulfide MSCs at different Cd oleate concentrations**
For the 500 mM and 100 mM Cd oleate preparation the only changes, from the 1000-mM method above, are the (1) amount of cadmium oxide added to 10.0 mL of oleic acid (0.64 g and 0.128 g of CdO, respectively), and (2) the amount of 2500 mM TOP=S added to achieved a 2:1 Cd:S ratio, reduced by a factor or 2 and 10, respectively.

For the 1580 mM cadmium oleate, 1.02 g of CdO was mixed with 5 mL oleic acid, and heated to 160 °C. At 150 °C, the formed cadmium oleate was a redish viscous paste (potentially with some undissolved CdO). Upon pulling vacuum at 90 °C, the solution foamed, so the temperature was gradually increased to degas the solution. The solution was cooled to 100 °C (below this temperature the solution begins to solidify), and 1.58 mL of 2500 TOP=S was injected, and the solution was heated to 130 °C.

**Different methods to prepare dilute 100 mM Cd oleate**
*Prepared 1000 mM cadmium oleate diluted to 100 mM with ODE.* The 1000 mM cadmium oleate was prepared as described above. After degassing and cooling to 50 °C. 1 mL of the 1000 mM cadmium oleate was mixed with 9 mL of ODE. After the injection of 2500 TOP=S (2:1 Cd:S ratio), the solution was heated to 140 °C before a color/turbidity change was observed, which is indicative of cluster/nanoparticle formation.

*Prepared 100 mM cadmium oleate with fixed acid/cd ratio.* The preparation is similar to 1000 mM Cd oleate, except only 0.128 g of CdO was mixed with 1 mL OA, to match the ratio in the 1000 mM preparation(~3.2:1 OA:Cd), and the balance of the solution was ODE (9 mL). The reaction was heated to 160 °C before a color/turbidity change was observed, which is indicative of cluster/nanoparticle formation.

*Prepared 100 mM cadmium oleate diluted with OA.* The preparation is similar to 1000 mM cadmium oleate, except only 0.128 g of CdO was mixed with 10 mL OA. The reaction was heated to 200 °C before a color/turbidity change was observed, which is indicative of cluster/nanoparticle formation.

**MSC synthesis with other carboxylic acid ligands**
The preparation of different cadmium carboxylates is similar to method for 1000 mM cadmium oleate described above. To match condition for the Cd oleate, the same molar ratio of acid:Cd was used (3:1) (see details in table below). The Cd to S ratio was still 2:1, with 2500 mM TOP=S as the sulfur source. Notably, the saturated chain metal carboxylates have higher melting points than unsaturated metal carboxylate (*i.e.*, Cd oleate). Thus, the injection temperature was increased until the solution was viscous, but not solid, specifically to 125, and 120 °C for the Cd laurate and Cd stearate samples, respectively.

**Table S1:** Summary of different Cd carboxylate preparation conditions.

| Acid | CdO (g) | Acid (g) | Acid (mL) | Acid/Cd molar ratio | Conc (mM) |
|---|---|---|---|---|---|
| Lauric acid | 1.075 | 5.04 | 5.70 | 3.00 | 1500 |
| Stearic acid | 0.708 | 4.7 | 5.57 | 3.00 | 1000 |
| Erucic acid | 0.543 | 4.3 | 5.00 | 2.99 | 850 |

**MSC synthesis with different solvents**

The 1000 mM Cd oleate solution was washed after degassing to remove the free oleic acid. To wash the Cd oleate, 1:2 parts toluene was added to the Cd oleate, and then ACN was added at 1:1 by volume. The solution was centrifuged at 4400 rpm for 5 min. This process was repeated 2 times, and then the Cd oleate was vacuum dried. The dried Cd oleate was mixed with a different solvent to achieve 1000 mM concentration Cd oleate with the different solvents, and the solution was heated to mix the two species (60 °C). Then, the required amount of TOP=S was added (2:1 Cd:S ratio). When cooled to room temperature, the mixture re-solidifies.

**Table S2:** Summary of preparation conditions for different solvents mixed with Cd oleate.

| Solvent | $Cd(OA)_2$ (g) | $Cd(OA)_2$ (mL)[a] | Solvent (g) | Solvent (mL) | 2500 M TOP=S (mL) |
|---|---|---|---|---|---|
| DDA | 0.477 | 0.394 | 0.259 | 0.312 | 0.141 |
| OLA | 0.458 | 0.379 | 0.244 | 0.300 | 0.136 |
| ODE | 0.452 | 0.373 | 0.233 | 0.296 | 0.134 |
| TOP=O | 0.448 | 0.370 | 0.258 | 0.293 | 0.133 |

[a] Cd Oleate density is ~1.21 g/mL (estimated as identical to cadmium stearate).

# Characterization Methods

**Absorption Spectroscopy**

Absorbance measurements were performed using an Ocean Optic USB 2000+ UV-Vis spectrometer with DH-2000-BAL light source. Before each sample, the appropriate solvent reference spectrum (hexane or empty cuvette) was measured. The samples were analyzed as-prepared (or without cleaning) unless otherwise noted.

For the ex-situ absorption experiment (**Figure S1**), the samples were analyzed as-prepared. Aliquots were taken every 15 min, and diluted by known factors until absorbance was <1. Notably, at longer times, the 1000 mM concentration reaction aliquots form a viscous gel in hexane, requiring extra-care to avoid mass loss during dilution.

**High Resolution Scanning Transmission Electron Microscopy (STEM)**

Lower magnification STEM images were acquired using an aberration-corrected FEI Titan Themis operating at 300kV, with a convergence semi-angle of 30mrad, and inner and outer collection angles of 68 and 340 mrad on an annular dark-field detector (**Figure S3** and **Figure** 2). The samples were prepared by drop-casting a solution of MSCs in hexane on a standard carbon coated Cu-TEM grid. The grids were then placed on a hot plate at 60 °C, and under vacuum to ensure that solvent was removed prior to imaging. Notably, the sample degrades after 7 s of plasma cleaning.

In effort to obtain atomic resolution images, lower MSC concentrations were used to isolate MSC assemblies, lower accelerating voltage (120 kV) and short exposure (2-5 s) were used to minimize beam damage, and cryogenic cooling was used to minimize carbon contamination. The convergence and collection angles were the same as in lower magnification imaging. Nevertheless, the MSCs still restructure and degrade quickly upon irradiation. Preliminary images of the clusters show that the MSCs are anisotropic, ~2 nm in size, and atomically disordered (**Figure S4**). Notably, beam damage prevented more thorough analysis. These difficulties (small size and beam damage) also impaired previous efforts to characterize clusters using TEM.[2–4]

**Small and Wide Angle X-ray Scattering (SAXS and WAXS)**
SAXS and WAXS measurements were performed simultaneously at A1 beamline at the Cornell High Energy Synchrotron Source (CHESS) using monochromatic radiation (wavelength = 0.62054 Å and bandwidth $\Delta\lambda/\lambda$ = 0.1%). The SAXS/WAXS images were collected by a ADSC Quantum-210 CCD detector with pixel size of 51.2 by 51.2 μm and total area of 4096 by 4096 pixels. The sample to detector distance was 630.824 mm, as determined from a silver behenate powder standard. Typical exposure time was 60 s. Images were dark current and geometric corrected, and then integrated using Fit2D software.[5,6]

The samples were heated in a custom designed apparatus (**Figure S6**). The sample holder was 2-mm thick aluminum sheet (25 by 30 mm) with a 13-mm hole in the center. The aluminum sheet was sandwiched on each side by 0.5-mm teflon sheet, and then kapton-tape (1 mil) covered washer (1-mm) with 6 2.5-mm holes drilled into each washer. The kapton tape was next to the Teflon, with the adhesive on the washer side. The sample was loaded into the sample holder, and holder was sealed with 6 bolts/nuts. Then, the sample holder was placed between two aluminum pillars that each held a heat cartridge. The temperature was controlled using a thermocouple within one of the aluminum pillar, near the heat cartridge. The sample temperature was monitored with a thermocouple placed inside the sample holder, and pushed to the bottom of the sample chamber. A 10-cm diameter (3.175 mm thick) Al disc was placed over the low-angle region of the detector to prevent saturation during 60 s scan (**Figures S6**). The beam passed through the sample near the bottom of the sample chamber.

For the ramp studies, the temperature was equilibrated at 100 °C, and then ramped to 170-200 °C at ~3-5 °C/min. The scattering for the empty cell at 130 °C was used for background subtraction, which introduces some error into the peak intensities because of the change in temperature; nevertheless, the observed peak positions and trends are accurate.

*X-ray Data Processing Method.* Data processing of integrated X-ray data included normalization, background subtraction, and re-scaling for the Al-disc at low-angles. After images were integrated, the curves were normalized to the incident beam intensity (**Figure S8a,b**), and the pattern for an empty cell was subtracted from the sample curve as a background. Notably because of differences in sample concentration, the beam was attenuated differently by the different concentration sample. To correct for this effect, the fraction of empty cell curve that was subtracted was adjusted to avoid subtraction artifacts at low-angles or low Q (**Figure S8c,d**). The fraction of the background curve intensity that was subtracted from each of the three concentrations tested pure oleic acid, 100, 500, and 1000 mM are 0.92, 0.89, 0.72, and 0.45, respectively. For the solid sample and 100 mg/mL resuspended in ODE, the fraction was 0.2 and 0.72, respectively. The partial subtraction does not significantly affect the peak position or intensity, and mainly remove low-intensity broad peaks due to the sample holder (**Figure S8**). (Note: for oleic acid the fraction was 0.92.) The low angle data (Q = $1.2 \times 10^{-3} - 7.5 \times 10^{-1}$ Å$^{-1}$) was also scaled by a factor of 15.28 (density = 2.7 g/cm$^3$, thickness = 3175 µm, Energy = 20 keV) to correct for the placement of the 10-cm diameter Al-disc used to attenuate the intense low-angle (SAXS) peaks on the detector.[7] To facilitate comparison, the SAXS/WAXS data was normalized to a large organic peak (~1.34 Å$^{-1}$) at the lower range of the WAXS (for example, see **Figure S7b**; the peak is more intense at lower Cd-oleate concentrations).

To quantify the loss of Cd oleate and the formation of the MSC assembly, the peaks were separate using the following method. The base of each hexagonal mesophase peak was linearly connected to create a continuous Cd-oleate curve. Then, both the Cd oleate and first mesophase peaks were integrated over the following range from 0.165 to 0.391 Å$^{-1}$ for Cd oleate and from 0.165 to 0.227 Å$^{-1}$ for the assembly (see **Figure S9** for an example).

**Nuclear Magnetic Resonance Spectroscopy (NMR)**
NMR spectra were acquired at 130 °C on a 500 MHz Bruker AVIII HD spectrometer equipped with a broadband Prodigy Cryoprobe. $^1$H spectra were acquired at 499.76 MHz with 8 transients, 45° (6 us) excitation pulse, 3.3 s acquisition time and 1 s relaxation delay. $^{113}$Cd spectra were acquired without decoupling at 110.86 MHz with 256 transients, 90° (12 us) excitation pulse, 87 ms acquisition time and 0.5 s relaxation delay. $^{13}$C spectra were acquired at 125.68 MHz with 128 transients, 30° (3.3 us) excitation pulse, 1.5 s acquisition time, 2 s relaxation delay and broadband $^1$H decoupling. $^{31}$P spectra were acquired at 202.30 MHz with 64 transients, 45° (6 us) excitation pulse, 0.4 s acquisition time, 2 s relaxation delay and broadband 1H decoupling. For *in-situ* reaction monitoring experiments acquisition of individual nuclei were interleaved in the order: $^1$H, $^{113}$Cd, $^{31}$P, and $^{13}$C. The total time for each cycle was 18 min and the total time was 10 h. Spectra were processed and analyzed with MNova 11.1 (Mestrelab Research S.L., Santiago de Compostela, Spain).

To prepare the reaction sample for NMR, the mixed solution of 1000 mM Cd oleate and 2500 mM TOP=S at 50 °C was loaded into to a NMR tube. The septum-capped and loaded NMR tube was placed under vacuum until bubbling stopped to remove air and moisture, and was then filled with nitrogen.

**Polarization Anisotropy Measurements**
The polarization anisotropy measurements along with accompanying absorption and emission (PL) spectra were performed using an absorption spectrophotometer (Jasco, V770 - UV/Vis/NIR) and fluorescence spectrophotometer (Edinburgh instruments, FL920). All measurements were carried out at room temperature. For PL, the sample is excited by the UV emission of a xenon lamp. The excitation and emission are spectrally filtered through double monochromator to improve signal to noise. Emission is collected by a fast PMT in a single photon counting scheme (SPC). For the anisotropy experiment, solution polarization PL was measured in photoselection scheme. The sample is excited with a vertically polarized light, which preferably excite particles aligned at this orientation. The emission is measured at vertical ($I_{VV}$) and horizontal ($I_{VH}$) polarization and the anisotropy (*r*) is calculated according to $r = \frac{I_{VV} - I_{VH}}{I_{VV} + 2I_{VH}}$. Anisotropy is a combination of the absorption and emission polarization (see ref.[8] for additional details on the polarization anisotropy). The clusters were suspended in hexane for the measurements.

**Time-resolved Fluorescence Lifetime**
Time-resolved fluorescence lifetimes (LT) were measured using time correlated single photon counting (TCSPC) scheme. The sample was excited by a 270 nm picosecond pulsed LED and measured with a fast photomultiplier tube (PMT). The total instrument response function was approximately 0.5 ns. The results were fitted by multiexponential decay (by deconvolution fitting), and an effective decay constant was defined by the time the fitted emission took to drop to 1/e of its maxima.

**Dynamic Light Scattering (DLS)**
Dynamic light scattering was performed on Zetasizer Nano-ZS (Malvern Instruments Inc). Sample solutions were prepared similarly to those used for TEM analysis. The cleaned and dried product was resuspended in hexane or ODE at 100 mg/mL, and then diluted to the order of 1-10 mg/mL. The 100 mg/mL was highly viscous and gel-like. The 1 and 5 mg/mL concentration solutions used for DLS were still noticeable more viscous than the neat solvent. A 2 min equilibration time was used for each measurement and three replicates were taken. The following solvent parameters were used hexane: refractive index = 1.375 and viscosity = 0.275 cP. The parameters for ODE were refractive index = 1.375 and viscosity = 2.84 cP.

# Figures and Tables

## Ex-situ Absorbance

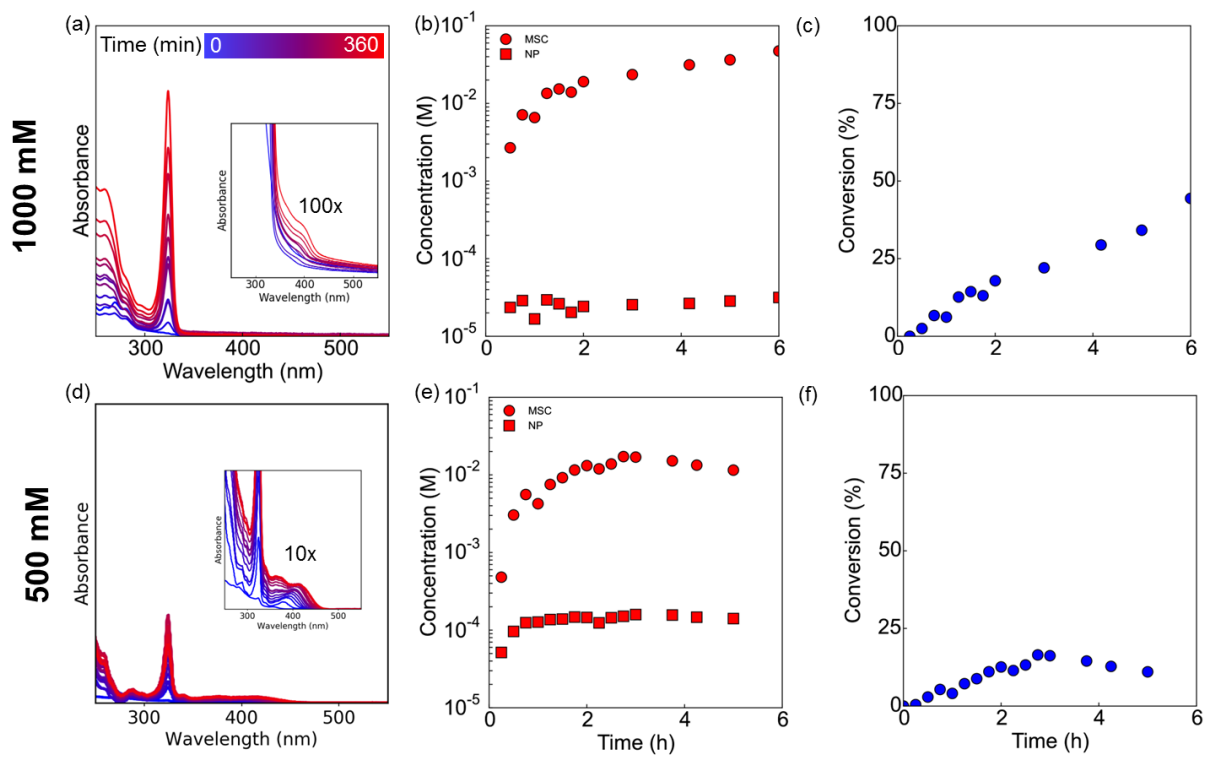

**Figure S1**: *1000 mM (a-c) and 500 mM (d-f) syntheses ex situ absorbance at* 130 °C. The absorption is for as-synthesized (uncleaned) samples. The MSC and NP concentrations and MSC conversion were calculated. (a) Absorbance shows the evolution of MSCs (excitonic peak at 324-nm) with time along with a much less intense (by factor of 100-1000) NP peak (Inset; absorbance value times by 100 compared to (a)). The size based on semi-empirical sizing curve is 1.6 nm for the clusters and increase from 2.7 to 3.4 nm for the nanoparticles (which corresponds to peak positions from 375 to 404 nm).[9] (b) Evolution of MSCs and NPs in terms of concentration (see calculation below), based on dilution factors and empirical extinction coefficient curve.[9] Notably, the MSCs concentration increases with time while NP concentration is nearly constant. At 6 h, the selectivity for MSC/NP is 1500, or MSC purity is 99.9%. (c) Precursor conversion into cadmium sulfide MSCs, increases linearly with time (with a Cd:S ratio[1] of 2:1; see calculation below). (d) For the 500 mM reaction, the absorbance shows the formation MSCs with a more intense NP peak compared to that for the 1000 mM reaction. (Note: (d) has the same y-scale as (a).) The NP peak is 10-100 times smaller than MSC peak (Inset: absorbance values times by 10 compared to (d)). The NP peak shifts from 358 to 418, corresponding to 2.2 to 4.0 nm NPs using a semi-empirical sizing curve.[9] (e) Evolution of MSCs and NPs in terms of concentration. Notably, the MSCs concentration increases with time, but then begins to decline after 3 h. At 3 h, the highest selectivity for MSC/NP is achieved (115), which corresponds to a MSC purity of 99.1%. (f) Precursor conversion into MSCs is lower for the 500 mM reaction compared to the 1000 mM, and at long times the MSCs decay, and NPs continue to grow.

## Scherrer Analysis (SAXS/WAXS)

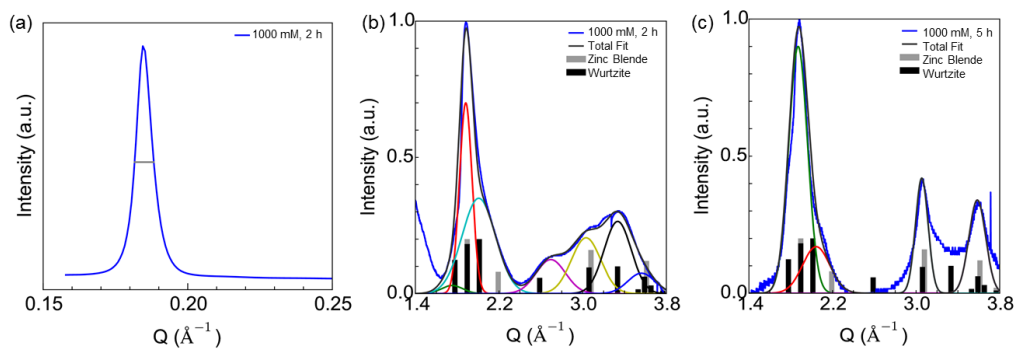

**Figure S2**: *Grain Analysis of 1000 mM Reaction at 130 °C (SAXS/WAXS).* (a) First hexagonal mesophase peak (SAXS) at 2 h (looks nearly identical at 5 h). Based on Scherrer analysis, the mesophase grain size is ~170 nm (86% of the peak width is from instrument broadening; see calculations). (b) CdS diffraction peaks (WAXS) at 2 h fit to wurtzite bulk peaks (see **Table S3** for details) correspondes to a 2.2±0.9 grain size. (c) CdS diffraction peaks (WAXS) at 5 h subtracted from diffraction curve at 2 h, and fit to zinc blende bulk peaks (see **Table S3** for details), suggests a 3.0±1.0 grain size.

*Table S3*: Summary of Peak Fitting for **Figure S2**

| Q (Å$^{-1}$) | Peak Position (°) | Intensity | Std Dev. (Å$^{-1}$) |
|---|---|---|---|
| **1.75** | 24.81 | 0.030 | 0.10 |
| **1.88** | 26.70[a] | 0.700 | 0.06 |
| **2.01** | 28.50[b] | 0.350 | 0.16 |
| **2.70** | 38.62[c] | 0.125 | 0.14 |
| **3.03** | 43.68 | 0.205 | 0.14 |
| **3.33** | 48.30[d] | 0.265 | 0.13 |
| **3.56** | 51.83 | 0.075 | 0.13 |
| **Scherrer Size = 2.2±0.9** | | | |

**1000 mM 5 h – Zinc Blende**

| Q (Å$^{-1}$) | Peak Position (°) | Intensity | Std Dev. (Å$^{-1}$) |
|---|---|---|---|
| **1.87** | 26.51 | 0.90 | 0.09 |
| **2.04** | 29.00[e] | 0.17 | 0.13 |
| **3.05** | 43.96 | 0.42 | 0.06 |
| **3.59** | 52.13 | 0.34 | 0.08 |
| **Scherrer Size = 3.0±1.0** | | | |

Shifted from bulk 2θ peak based on Cu source. Bulk peak positions are [a]26.51°, [b]28.18° [c]36.62°, [d]47.84°, and [e]30.81°.

## STEM images

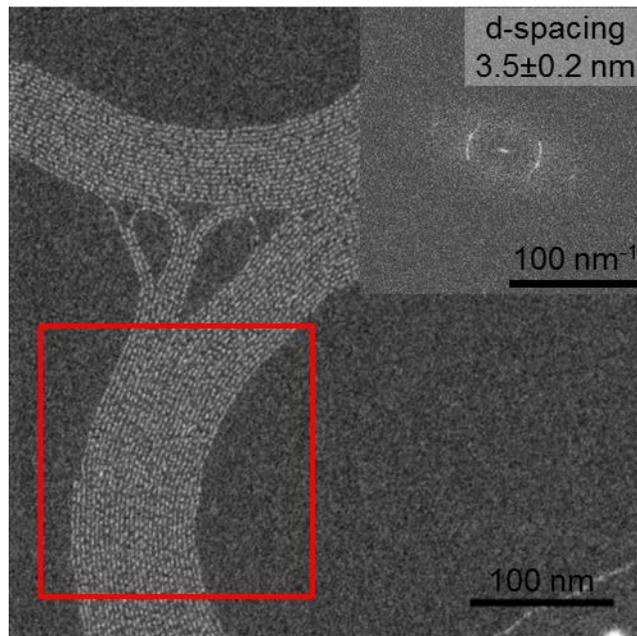

**Figure S3**: *STEM fibers spacing.* STEM image of cleaned high-purity MSCs (1000 mM reaction). Inset: FFT of red-boxed region illustrating the 3.5 nm spacing of the nanofibers. Notably, the d-spacing from the SAXS is similar with a d-spacing of 3.4 nm.

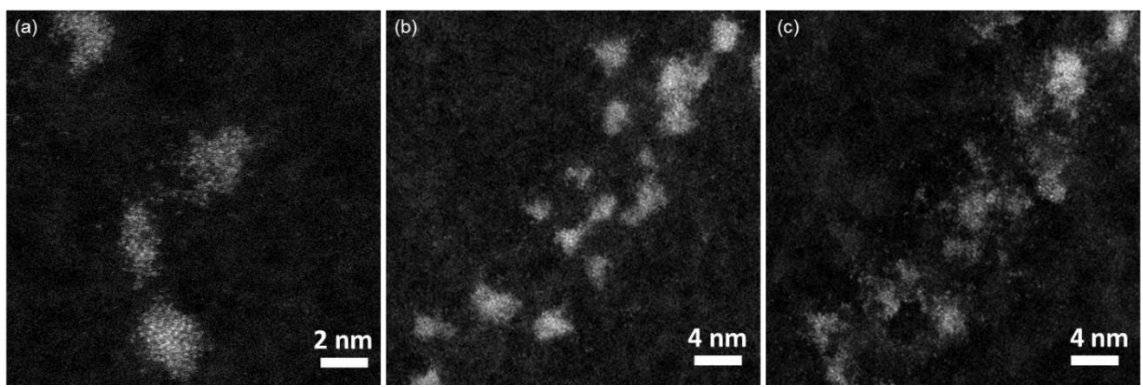

**Figure S4**: *High resolution STEM of MSCs.* (a) Rapidly acquired STEM images at high magnification, prior to additional irradiation (where beam damage is expected to be minimal), shows disorder size and shape. (b,c) Lower magnification images of clusters, before and after high magnification imaging (i.e., high electron dose), indicates that the MSCs degrade under irradiation doses. Scale bar are same physical length, despite difference in scale.

## Photoluminescence Spectroscopy

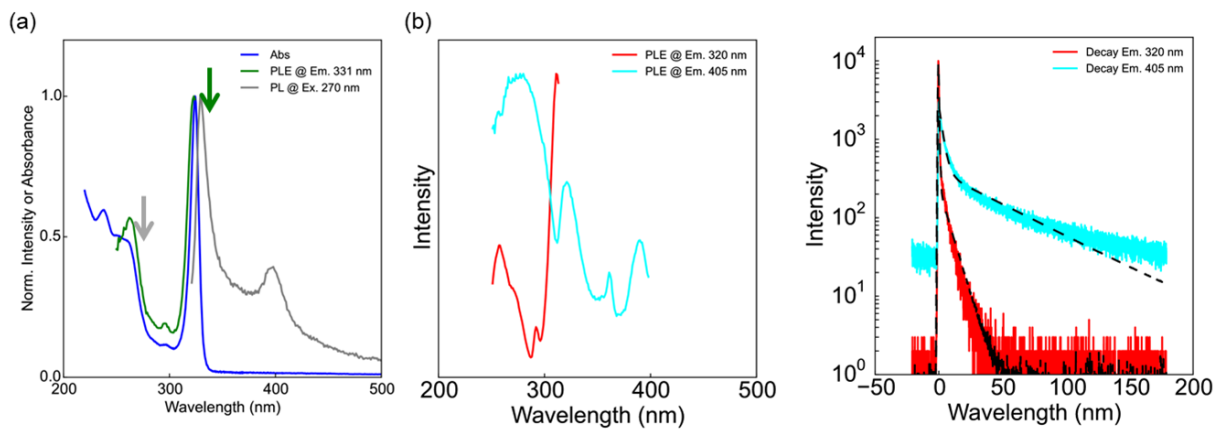

**Figure S5**: *Electronic Structure Characterization.* (a) Absorbance, PL excitation (PLE) and PL for 324-nm clusters in hexane. For PLE, the emission wavelength was at 331 nm (green arrow). For PL, the excitation wavelength was at 270 nm (grey arrow). (b) Comparision of PLE at emission of 320 and 405 nm. (c) Comparison of PL lifetime for 320 and 405 nm emission, which are 0.68 and 4.15 ns, respectively.

Apparatus

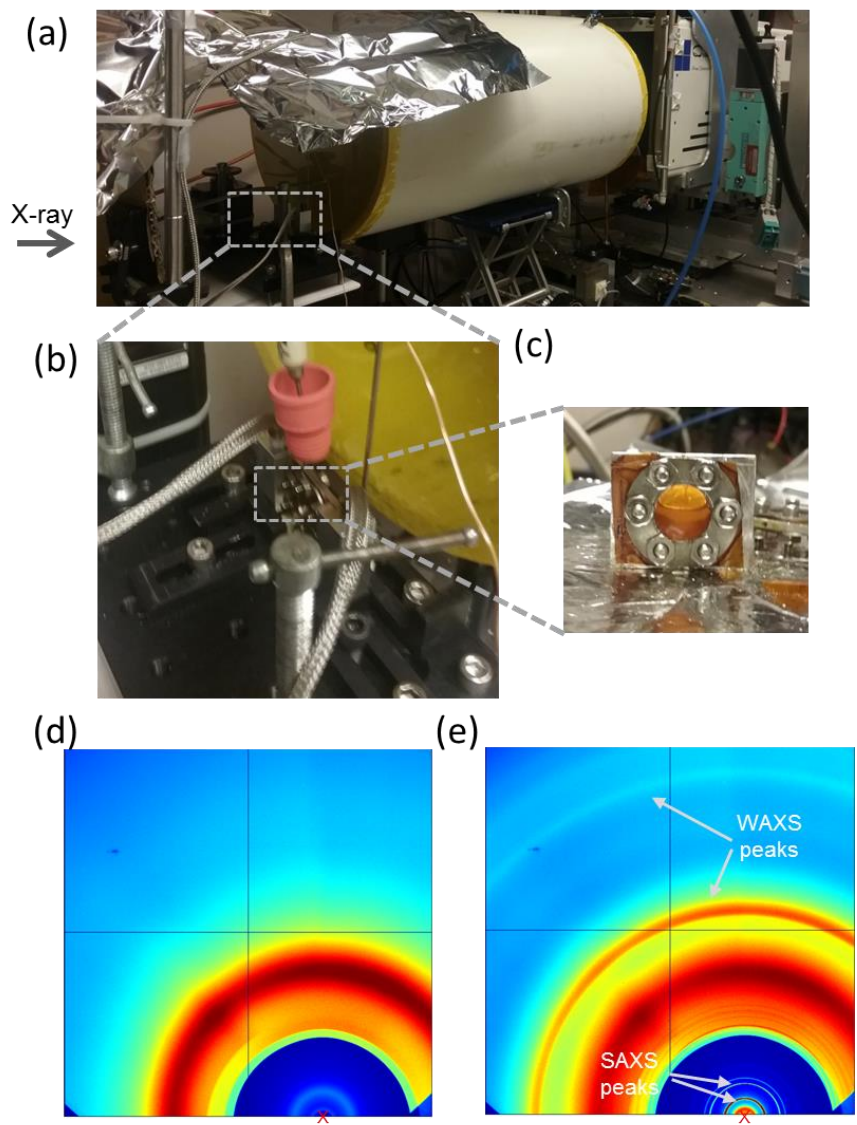

**Figure S6**: *In-situ SAXS/WAXS apparatus and 2D scattering images.* (a-c) Images of the apparatus, flight path, and detector. (b) Zoomed-in view of the apparatus and heating stage, and (c) sample holder with sample. (d) Image from the detector for 1000 mM reaction at 0 h at 130 °C. The dark blue region around the beam center (marked by red X) is due to attenuation of the Al-block. The faint blue ring near the beam center is the Cd oleate SAXS peak. The intense red peak relates to organic molecules. (e) 1000 mM reaction after ~5 h at 130°C. The intense and sharp SAXS rings are visible near the beam center, and propagate past the Al-block region. The WAXS region shows the formation of weak (light blue) diffraction rings.

X-ray Data Analysis

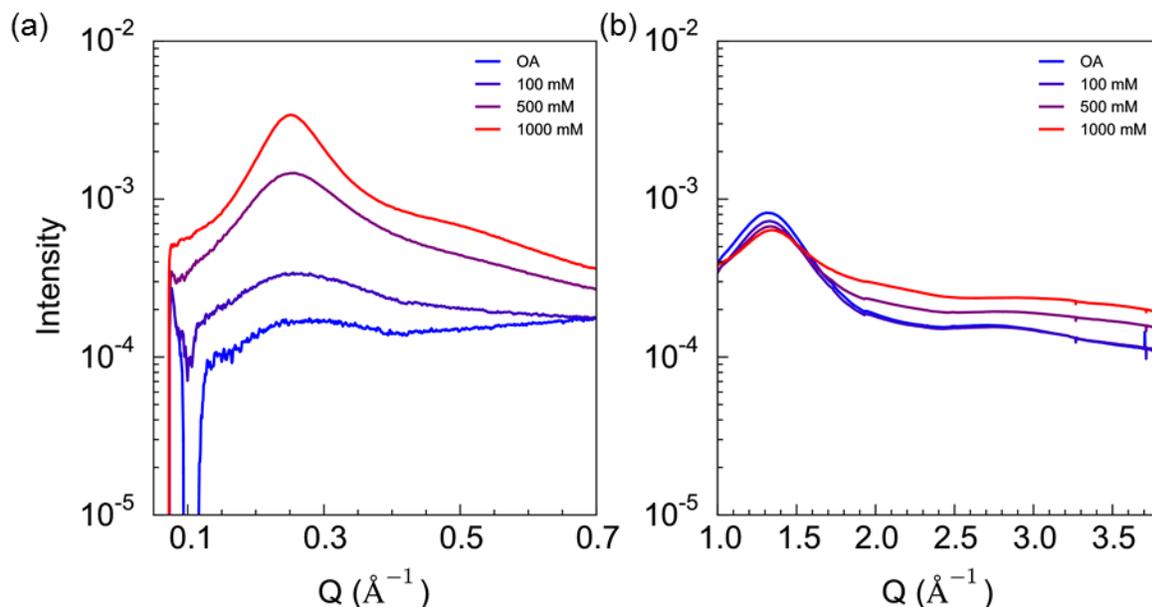

**Figure S7**: *SAXS/WAXS precursors at 130°C (background subtracted).* (a) SAXS pattern for the precursors shows a peak at ~0.251 Å$^{-1}$ (or 2.5-nm d-spacing) that increases with concentration, likely correspond to cadmium oleate micelles or coordination polymers. The FWHM of the 1000 mM is 0.12 Å$^{-1}$, corresponding to grain size of 10 nm. (b) the WAXS is relatively featureless, except for peak at ~1.2 Å$^{-1}$ , which is more intense for less concentrated samples. This peak likely corresponds to organic molecules, and is similar to data previous reported for octadecene[10] (See **Figure S8** and Methods section for discussion on background subtraction method)**.**

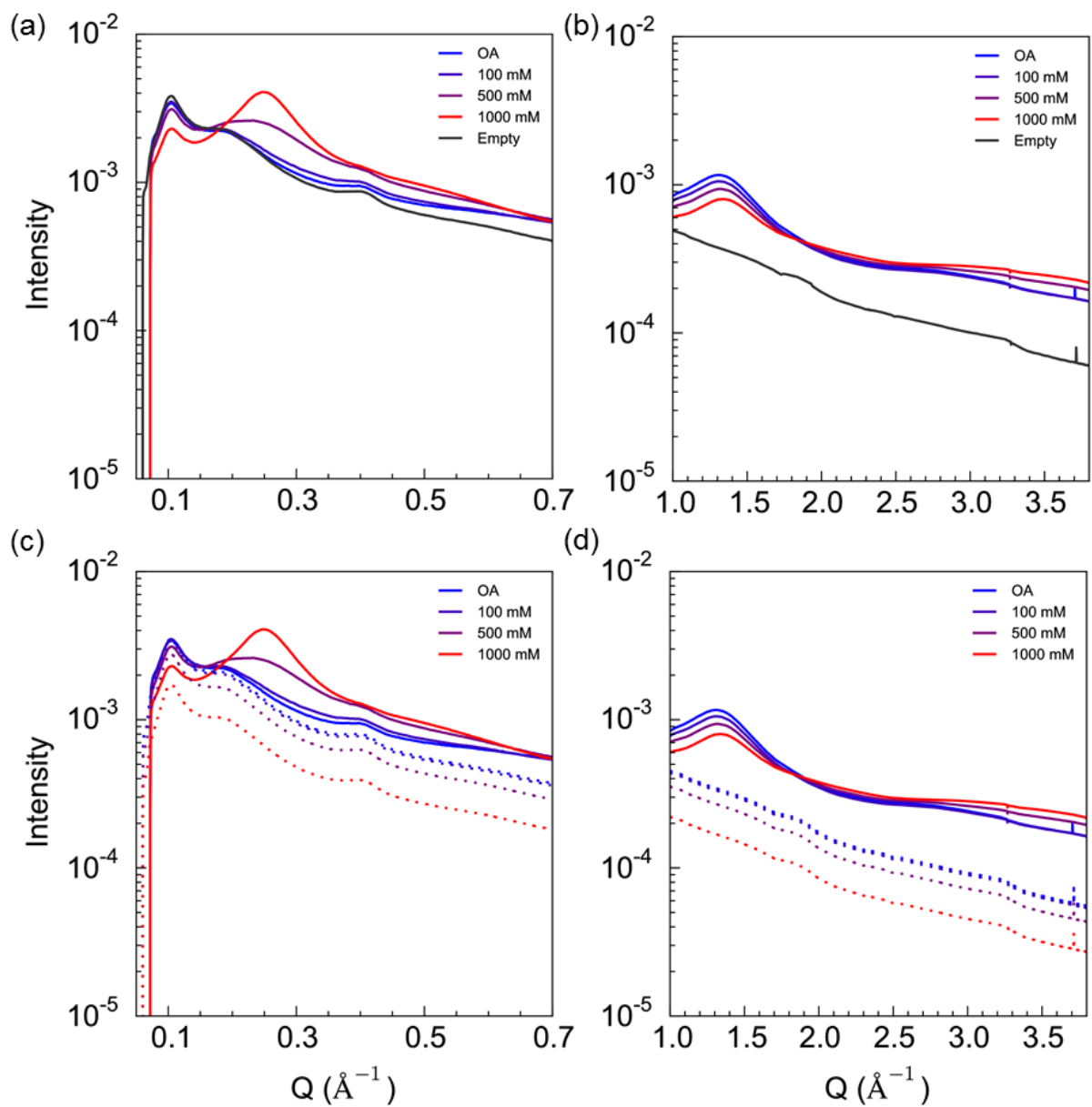

**Figure S8**: *Precursors at SAXS/WAXS at 130 °C.* (a,b) SAXS and WAXS data normalized for incident intensity, but without subtraction of empty cell (shown as "grey" curve). (c,d) Solid lines are same as (a,b), but the dashed lines are background contribution to solid line signal (color matched). Notably, the empty cell intensity at low-Q, <0.15 Å$^{-1}$, exceeds that of the sample; thus, only a percentage of total background was subtracted (see **Methods** for discussion).

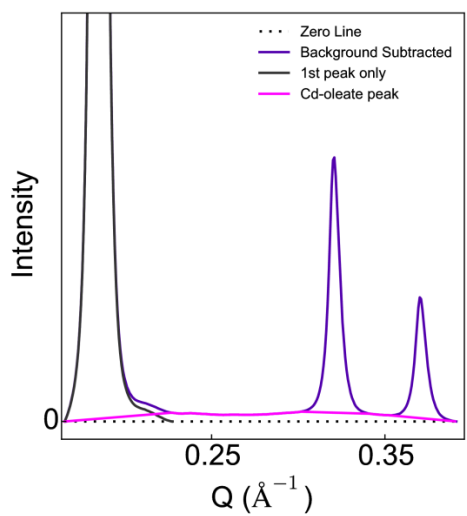

**Figure S9**: *Separating the MSC assembly and Cd oleate SAXS peaks.* To separate the first mesophase peak from the Cd oleate peak, the base of each hexagonal mesophase peak was linearly connected to form a close Cd oleate region for integration (pink curve above). The Cd oleate region was integrated from 0.165 to 0.391 Å$^{-1}$ while first hexagonal peak was integrated from 0.165 to 0.227 Å$^{-1}$. The data shown is after 98 min at 130 °C for the 1000 mM reaction.

## 1580 mM reaction (SAXS/WAXS)

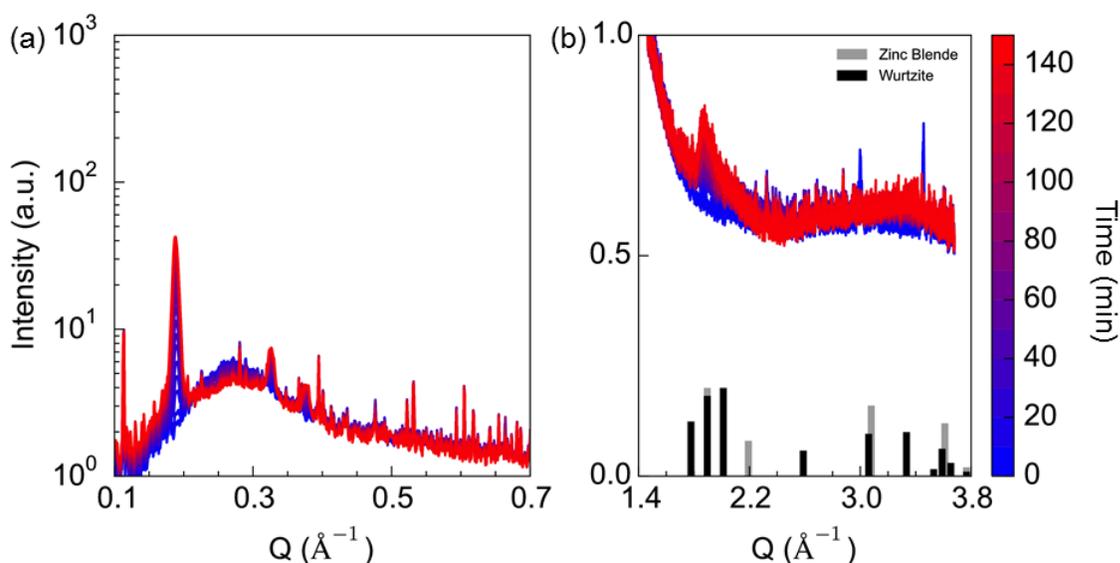

**Figure S10**: *1580 mM Reaction (SAXS/WAXS) at 130 °C*. (a) The hexagonal mesophase peaks (SAXS) are observed in the SAXS similar to 1000 mM reaction. (Note: both the SAXS and WAXS are much noisier that the other data sets because the detector cooling system shut off during the experiment. Nevertheless, the formation of hexagonal mesophase is still observed.) The first peak position is 0.1878 Å$^{-1}$ is shifted slightly from the 1000 mM reaction (0.1845 Å$^{-1}$). (b) CdS diffraction (WAXS) does not change significantly upon heating. Specifically, no sharp peaks are observed to indicate NP growth, suggesting that the MSCs do not grow via coalescence, even at temperature as high as 170 °C. The strong shoulder around 1.4 Å$^{-1}$ is due to ODE solvent (not its absence in **Figure S15**).

NP Structure Factor

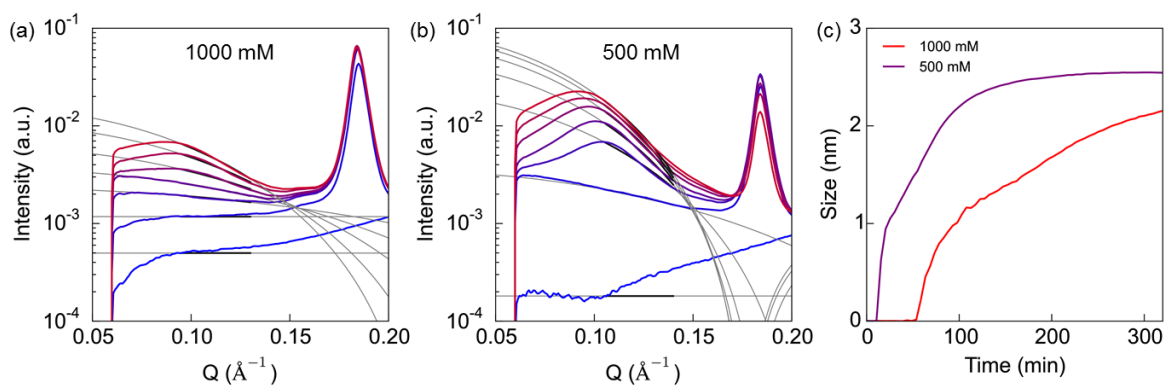

**Figure S11**: *NP structure factor (SAXS) for 500 and 1000 mM reactions.* SAXS patterns (colored) and corresponding least-squared fits (light grey is simulated fit; dark grey is fitting region, which was 0.095 to 0.13 Å$^{-1}$ for 1000 mM and 0.105 to 0.14 Å$^{-1}$ for 500 mM) to the monodisperse sphere model[11] for the (a) 1000 mM and (b) 500 mM samples. Selected traces and fits are shown (every 50 min from 0 to ~320 min (from blue to red curve)). The fits based on the form factor for monodisperse spheres and resulting sizes are a rough approximation for this system for three reasons: (1) the SAXS patterns represent the structure factor rather than the form factor because of the high concentrations used (as evidenced by the intensity decline at lower Q (0.05-0.1 Å$^{-1}$, whereas the modeled intensity increases); (2) this cursory fit does not include any particle size dispersion; and (3) the SAXS pattern is only measured over a narrow range (0.05-0.15 Å$^{-1}$), preventing more complete fitting/modeling of the structure factor. (c) Simulated particle size based the spherical form factor fits for all time traces (every 5 min). The 1000 mM has a delayed onset of particle growth and slower growth rate compared to the 500 mM reaction. At early times, the structure factor is dominated by the cadmium oleate (blue curves in (a) and (b)). The final size is on par with that estimated from the empirical sizing curve 1.6 nm for the MSCs and 2 to 4 nm for NPs (**Figure S1**).

1000 mM Higher Temperatures (SAXS/WAXS)

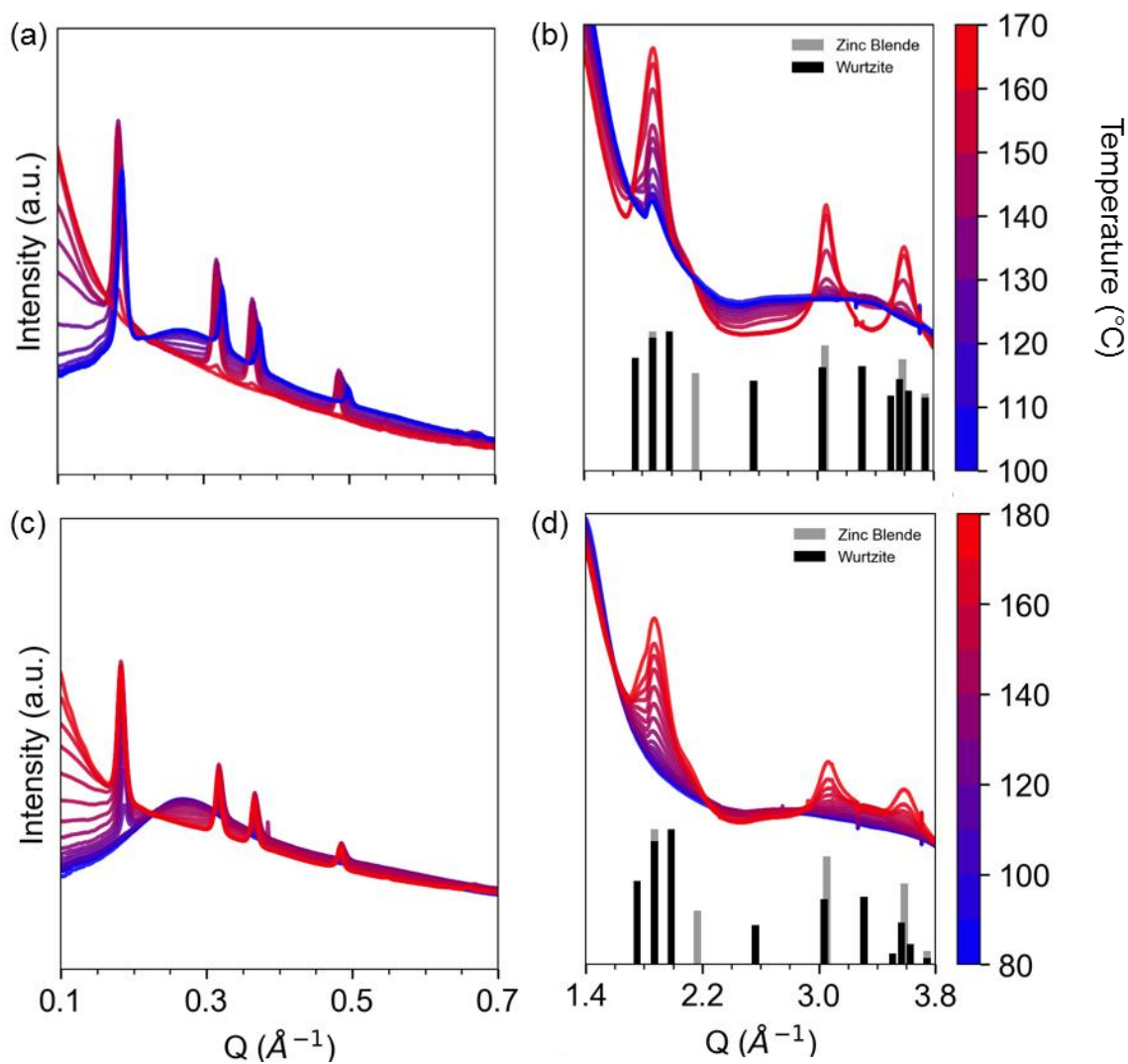

**Figure S12**: *1000 mM reaction at higher temperature (>130 °C).* (a-b) The 1000 mM Cd oleate + TOP=S solution was heated from 100 to 170 °C at 3 °C/min, and held at 170 °C for 35 min. Heating the solution to elevated temperatures led to the decay of the hexagaonal mesophase peaks (shown in (a)) and growth of zinc blende NP at the expense of the wurtzite-like MSCs (shown in (b)). NOTE: The solution was held at 100 °C for 1.5 h prior to the experiment due to beam stability issues, during this time the reactions slow preceded, as indicated by mesophase and CdS peak at 100 °C. (c-d) Same 1000 mM Cd oleate + TOP: S solution ramped from 80 to 170 °C at 3 3°C/min, but not held at 170 °C (or initially at 100 °C for a prolong period of time).

## Heating Resuspended MSCs (SAXS/WAXS)

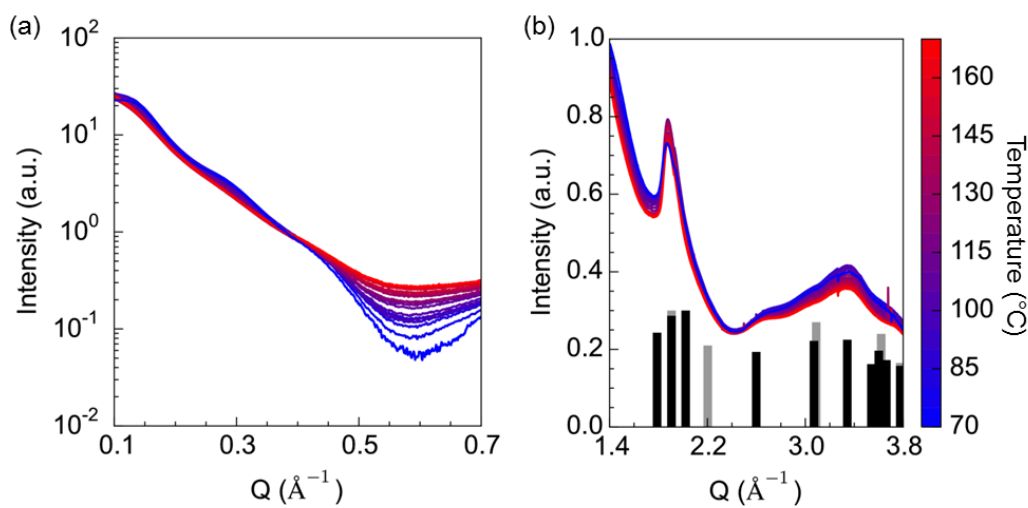

**Figure S13**: *Heating MSCs in ODE (100 mg/mL; SAXSWAXS).* Cleaned MSCs were resuspended in ODE, forming a viscous gel. The solution was heated from 70 to 170 °C at 3 °C/min ramp rate. (a) The mesophase peaks (SAXS) are not observed in the SAXS, but the viscous nature of the solution suggests it is still fibrous, potentially from unbundling of the mesophase. (b) CdS diffraction (WAXS) does not change significantly upon heating. Specifically, no sharp peaks are observed to indicate NP growth, suggesting that the MSCs do not grow via coalescence, even at temperature as high as 170 °C. The strong shoulder around 1.4 Å$^{-1}$ is due to ODE solvent (note its absence in **Figure S15**).

## Dynamic Light Scattering

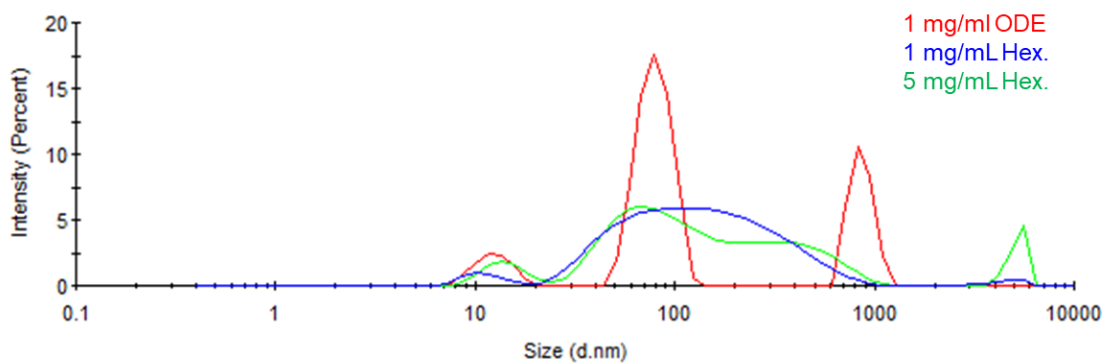

**Figure S14**: *Dynamic light scattering of resuspended MSCs.* Cleaned MSCs were resuspended in hexane (hex.) and ODE. The 1 mg/mL MSC in hexane (blue curve) have a good fit, and shows large aggregates (100's nm). The 5 mg/mL in hexane and 1 mg/mL in ODE samples gave poor fits due to large polydispersity and aggregation with time. Nevertheless, the light scattering indicates that large aggregates are present in cleaned and resuspended solution. These aggregates likely contribute to the viscous and gel-like behavior of the MSC solutions. NOTE: The aggregates sizes are approximates due to the increased solution viscosity relative to the neat solvent viscosity, and large polydispersity.

Heating Solid MSCs (SAXS/WAXS)

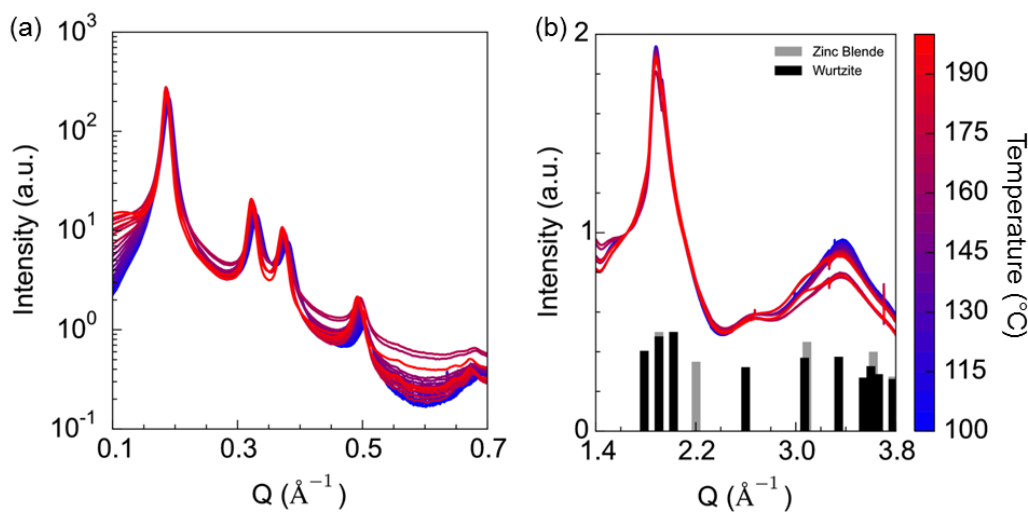

**Figure S15**: *Heating solid MSCs (SAXS/WAXS).* The solution was heated from 100 to 200 °C at 3 °C/min ramp rate. (a) The mesophase peaks (SAXS) is stable upon heating, and expand by ~3% in d-spacing over the 100 °C range. (b) CdS diffraction WAXS does not change significantly upon heating. Above 180 °C, there is a slight increase in a zinc blende peak (3.0 Å$^{-1}$). This result suggests that heating the solid (with minimal MSC mobility) does not lead to significant growth, but some growth likely via coalescence may be possible above 180 °C.

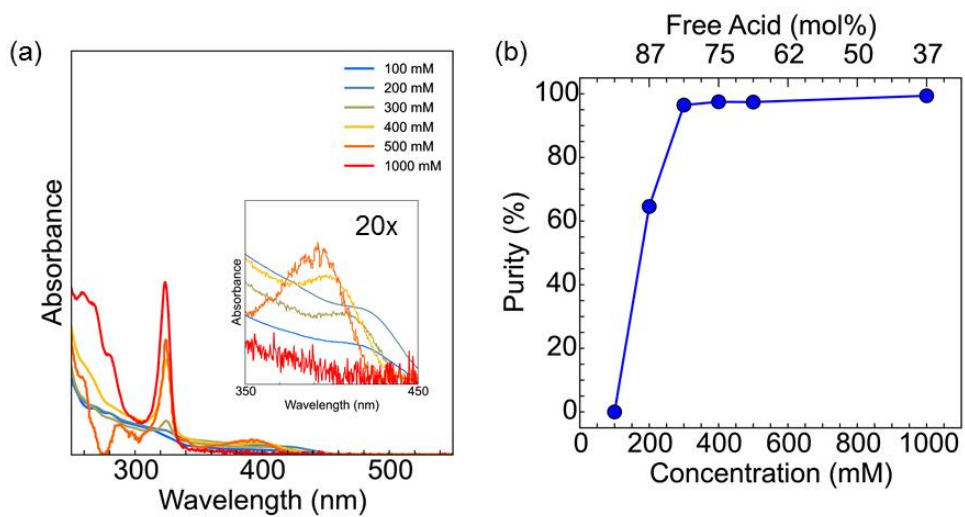

**Figure S16**: *MSC-to-NP Crossover.* (a) Absorbance for a series of reactions at 130 °C for 1 h at various concentrations. The 500 and 1000 mM are 1 h point from **Figure S1**. All concentrations ≥200 mM form MSCs. Inset: 20 times magnification of NP peaks. (b) Purity of MSCs as function of concentration. The free acid concentration is calculated as the amount of free acid over the total amount of free acid and oleate. The ratio of Acid/Cd is inversely proportional to concentration: Acid/Cd ratio= 3169/Conc[mM].

100 mM Preparation (Absorbance)

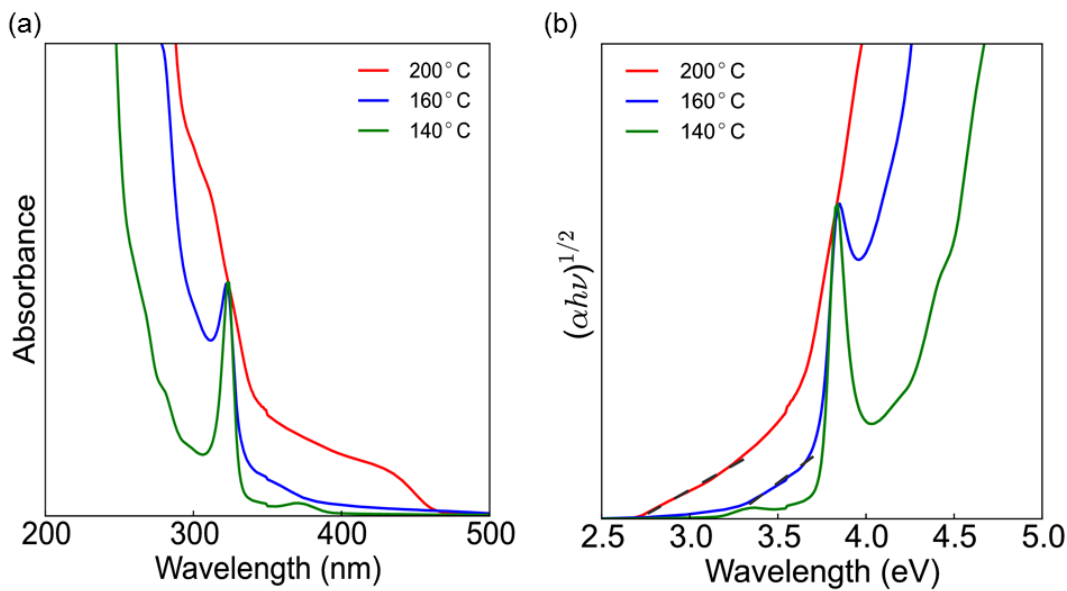

**Figure S17**: *100 mM preparation methods.* To dilute the 1000 mM solution, three different combinations are possible: (1) prepared concentrated and dilute in non-coordinating solvent (*i.e., ODE)* ("green" curve, 140 °C), (2) prepared dilute with fixed acid to Cd ratio (3:1 with balanced ODE) ("blue" curve, 160 °C), and (3) prepared dilute with pure oleic acid ("red" curve, 200 °C). The solution was heated until a color/turbidity change was observed. (a) Absorbance *vs.* wavelength. (b) Tauc plot to determine NP peak position. The NP peak is at 369, 383, 458 nm for conditions (1), (2), and (3), respectively (values in eV are 2.71, 3.23, and 3.36). The purity between MSCs and NPs was calculated using Peng et al.[9] sizing curve and extinction coefficient (see Calculations). The purity for the three conditions was 73, 80, and 0%, respectively. The sizes are 2.5, 2.9, and 5.6 nm, respectively. The results for case 2 and 3 are consistent with Peng et al. that demonstrated excess oleic acid causes Ostwald ripening based growth, whereas case 1 indicates that the concentrated cadmium oleate precursor creates an inherently different synthesis environment. This difference is likely due to the formation of cadmium oleate coordination polymers, which is favored by enhanced precursor-precursor interactions at high concentrations as compared to dilute conditions. All spectra are for as-synthesized (uncleaned) clusters.

The variety of ways to create a dilute (100 mM) solution highlights a key different between high concentration and dilute synthesis, namely the variety of different species interactions. High precursor concentrations reduce the variety of species in solution, promoting precursor-precursor interactions over precursor-solvent or –surfactant interactions. Specifically, 1000 mM cadmium oleate has an overall 3:1 acid to Cd ratio, or 1 mol cadmium oleate (*i.e.*, 2 mol acid per mol Cd) to 1 mol free oleic acid. At higher concentrations (1500 mM cadmium oleate, ~2:1 oleic acid:Cd), the solution is only cadmium oleate with no free solvent. The 2500 mM trioctylphosphine sulfide (TOP=S) is a 1:1 ratio of phosphine to sulfide, eliminating the effect of free TOP molecules. In contrast, dilute solutions have a wide variety of species in solution and less precursor-precursor interactions. Overall, the high concentrations refine the NP synthesis environment by maximizing the amount of precursor-precursor interactions while minimizing the amount precursor to solvent and/or free ligand interactions.

Precursor Schoichiometry

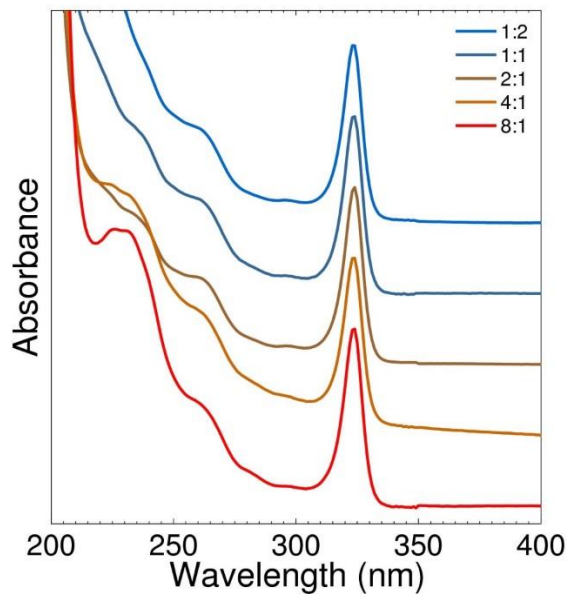

**Figure S18**: *Cd:S precursor stoichiometry.* As-synthesized (uncleaned) 1000 mM reaction at 140 °C with different precursor stoichiometric ratios, ranging in precursor stoichiometry from Cd:S 1:2 to 8:1. In all cases, 324-nm MSCs are formed. Note: these experiments were performed at 140 °C rather than 130 °C, increasing the temperature increases the production of MSCs and does change the selectivity (as long as temperature is below 150 °C at which point NP growth increases).

Different Ligand Lengths

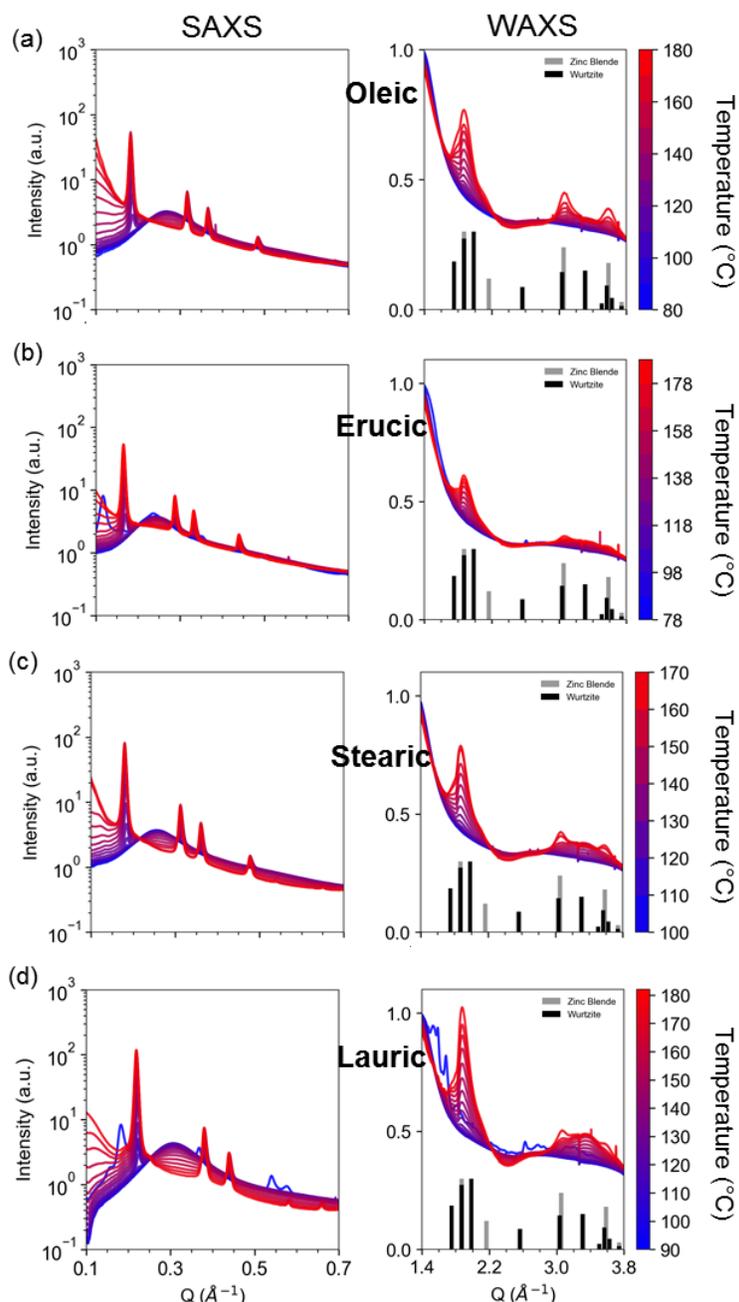

**Figure S19**: *Comparison of MSCs synthesis using different ligand lengths (SAXS/WAXS).* The solution was heated from ~80 to ~180 °C at 3 °C/min ramp rate. In all cases, the samples show a micellar peak, form a hexagonal mesophase (along with a wurtzite crystal phase) and show ZB peaks (3.0 Å$^{-1}$) at elevated temperature (150-180 °C). Notably, the erucic and lauric acid based Cd precursors had not melt by 80 °C, and show lamellar peaks (SAXS), that then fade into a single micellar peak at higher temperatures (>100 °C). The following carboxylic acids were investigated (a) oleic acid, (b) erucic acid, (c) stearic acid, and (d) lauric acid. **Figure S20** shows the evolution of the peak max peaks position and **Figure S21** shows a comparison between the four acids at 142 °C

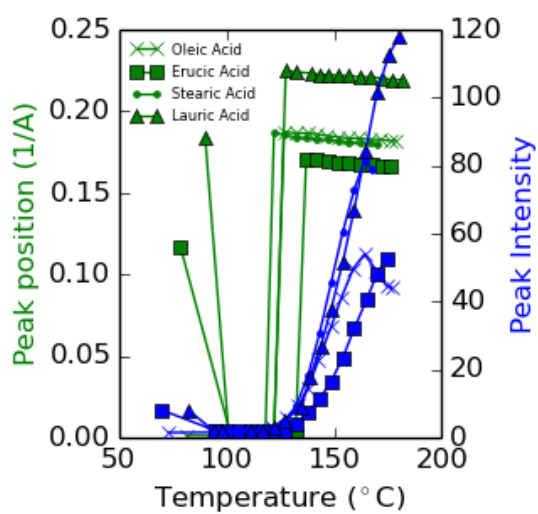

**Figure S20**: *Evolution of mesophase peak with temperature for different ligands.* A comparison of the maximum peak position and intensity within ~0.01A$^{-1}$ of hexagonal first peak (Note: peaks positions farther away than ~0.01 are considered to be zero; **Figure S18** and **Table S4**). At low temperatures, only the erucic and lauric acid samples show a peak position because their lamellar phase has not yet melted while the oleic and stearic acid have melted into a micelle structure. The formation temperature for the mesophase for all four samples is around 120-150 °C as indicated by the rapid increase in peak position (green curves), and increase in peak intensity (blue curves). Specifically, the oleic, erucic, stearic, and lauric acid samples form MSCs at 127, 137, 122, 151° C. At higher temperatures (above 150 °C), the stearic acid and oleic acid peak intensity begin to decrease, implying the fading away of the mesophase.

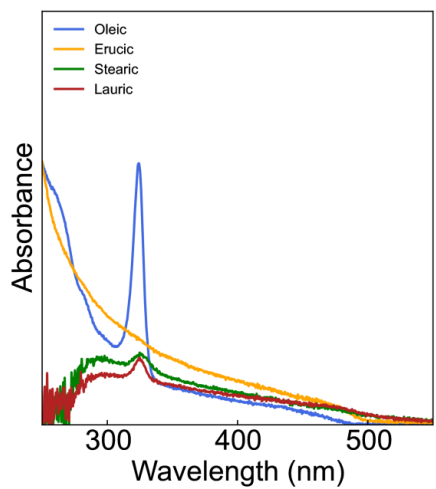

**Figure S21**: *Absorbance of final sample using different ligands (from **Figure S19**).* The oleic, stearic, and lauric acid based samples all show a 324 nm excitonic peak, and a weaker NP peak (~470 nm). The stearic and lauric sample did not resuspend well leading to noisy spectra. The erucic acid based sample does not show a cluster excitonic peak, but is expected to have formed clusters based on the mesophase (see **Figure S19**). We hypothesize that the lack of a cluster peak is due to subsequent reaction before the acquisition of the absorption spectra. All spectra are for as-synthesized clusters, without cleaning.

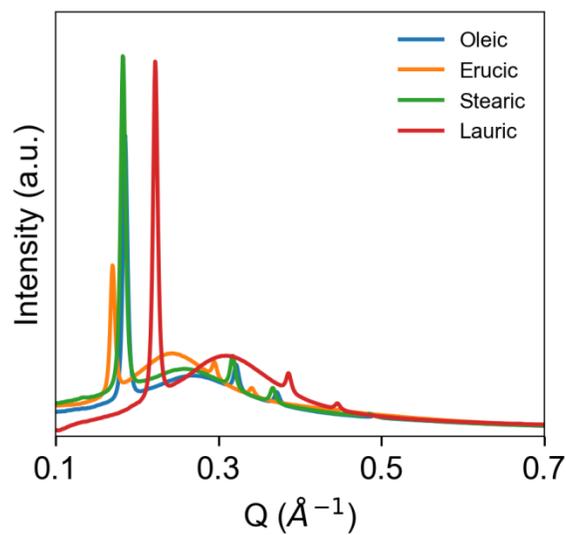

**Figure S22**: *Comparison of MSC mesophase made with different ligands at 142 °C.* Comparison of hexagonal mesophase peaks for MSCs synthesized with different carboxylic acid ligands at 142 °C (full temperature study shown in **Figure S19**). The oleic and stearic give similar d-spacings while the shorter lauric acid and longer erucic acid give shorter and longer d-spacings (or higher and lower Q values), respectively. Notably, d-spacing is much shorter than two ligand lengths suggesting that the ligands are likely interdigitated and bent (see **Figure S23** and **Table S4**).

**Table S4**: Comparison of mesophase d-spacing and ligand length for different carboxylic acids (**Figure S22**). The gap is the different between the d-spacing and the cluster diameter. The cluster diameter is 1.6 nm based on absorbance.[9] The ratio of the Gap/(2•Ligand) indicates that the gap is half the distance required for two ligands to be full-extended and not overlapping, indicating that the ligands are likely interdigitated and bent. Carbon # indicates the number of carbons in the chain, and = means there is a double bond in the chain. 1Q is the position of the first hexagonal peak. Both oleic and erucic have a double bond at the 9 position.

| Acid | Carbon # | 1Q (Å$^{-1}$) | d (nm) | Ligand Length (nm) | Gap (nm) | Gap/(2•Ligand) (%) |
|---|---|---|---|---|---|---|
| Oleic | C18, = | 0.1856 | 3.39 | 2.02 | 1.79 | 44% |
| Erucic | C22, = | 0.1701 | 3.70 | 2.49 | 2.09 | 42% |
| Stearic | C18 | 0.1831 | 3.43 | 2.32 | 1.83 | 39% |
| Lauric | C12 | 0.2226 | 2.82 | 1.57 | 1.22 | 39% |

S24

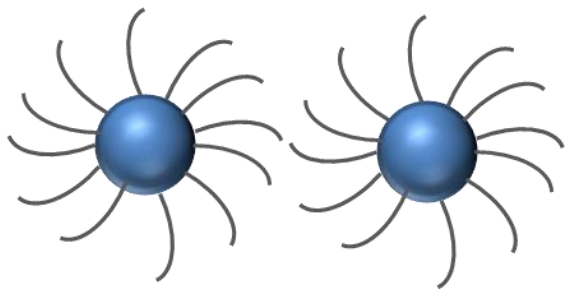

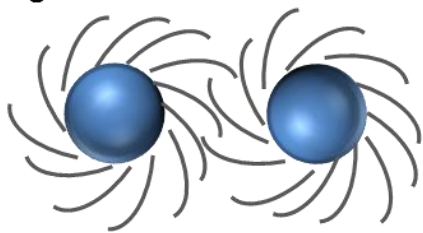

**Figure S23**: *Schematic of ligand configuration based on mesophase d-spacing.* The d-spacing of the mesophase is less than a two ligand lengths, indicating in contrast to the conventional view that the ligands are likely bend and interdigitated ("Emerging View"). **Table S4** provides calculated gap distance between the particles compared to the ligand lengths.

## 1000 mM Different Free Solvents

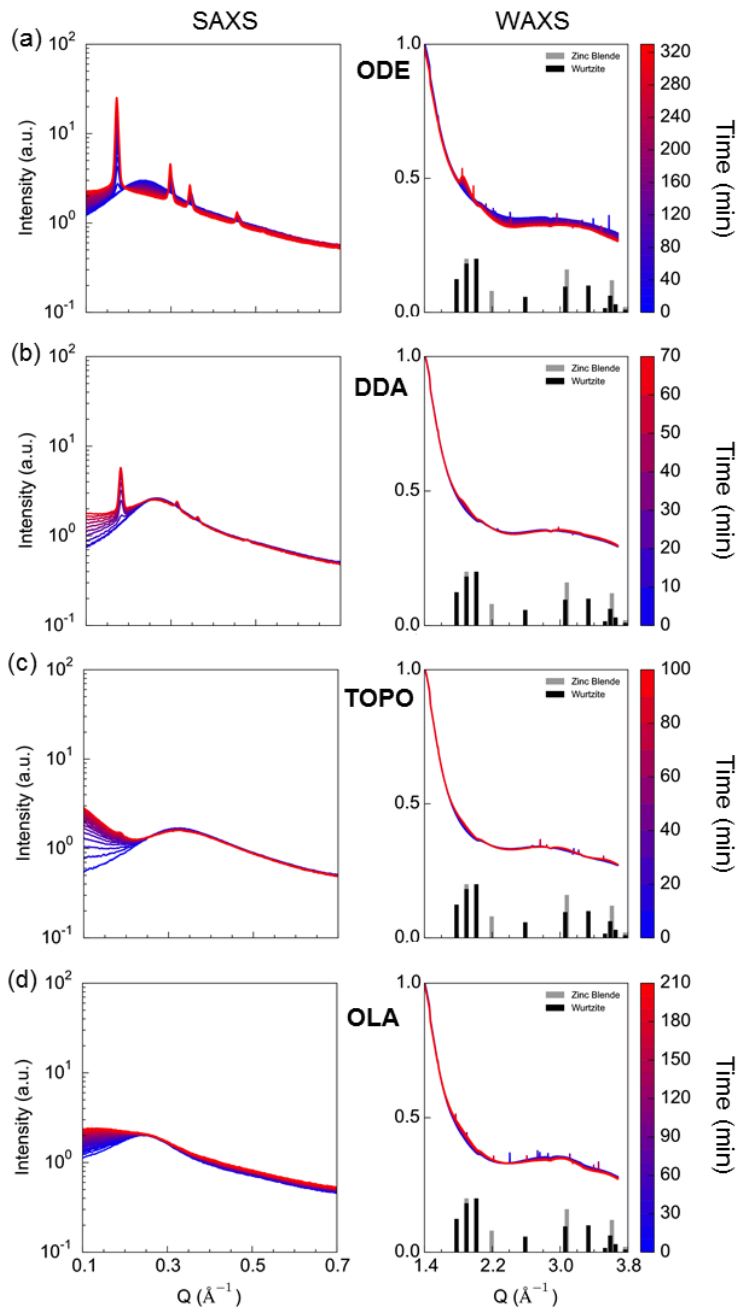

**Figure S24**: *MSC reaction with 1000 mM Cd oleate in various solvents at 130 °C (SAXS/WAXS).* Cleaned and dried Cd oleate (prepared as 1000 mM before cleaning) and then mixed with different solvents, rather than the oleic acid present during typical concentrated synthesis, namely three coordinating (DDA, TOP=O, OLA) and one non-coordinating (ODE). The concentration of the mixed solution was approx. 1000 mM Cd oleate. The precursor conversion is low in all four cases evidenced by the persistence of a strong Cd oleate peak (SAXS ~ 0.25 Å$^{-1}$). (a) SAXS/WAXS of mixture with ODE (a non-coordinating solvent) shows the formation of hexagonal mesophase and weak CdS diffraction peaks. (b) SAXS/WAXS mixture with DDA shows the formation of hexagonal mesophase and weak CdS

diffraction peaks. (c) SAXS/WAXS mixture with TOP=O displays the formation of hexagonal mesophase, albeit with low intensity, and weak CdS diffraction peaks. (d) SAXS/WAXS mixture with OLA does not show the formation of hexagonal mesophase, but does show weak CdS diffraction peaks. Note: The reaction times are not identical (a-d). Reaction times for ODE, DDA, TOP=O, and OLA samples are approx. 340, 80, 110, and 230 min, respectively.

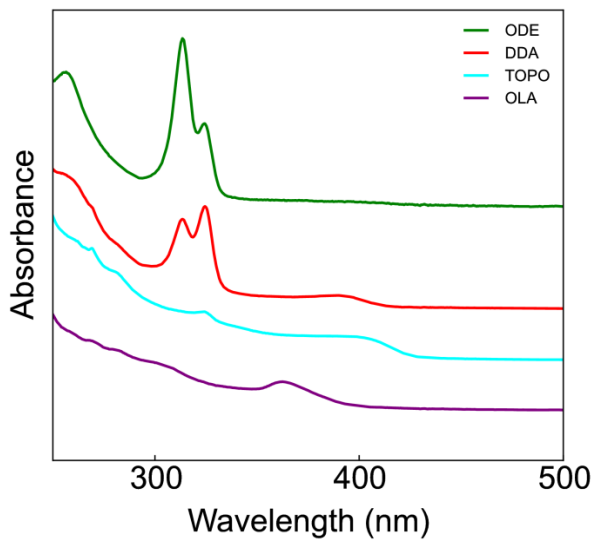

**Figure S25**: *MSC reaction with 1000 mM Cd oleate in various solvents at 130°C.* Absorbance of the final uncleaned sample. The ODE, DDA, and TOP=O show MSC clusters peaks at 313 and/or 324 nm. (Note: 313-nm MSC is the result of prolonged (~1-2 days) air-exposure on the F324, and is suspected to related to the absorption of moisture.) The DDA and TOP=O samples also have a significant NP peak at ~400 nm. The OLA sample shows a MSC peak at ~360 nm[1] with a shoulder to the right that may indicate NP growth. The reaction times were different between each sample, but illustrate the trends of more coordinating solvent promoting NP growth. Reaction times for ODE, DDA, TOP=O, and OLA samples are approx. 340, 80, 110, and 230 min, respectively. All spectra are for as-synthesized (uncleaned) samples.

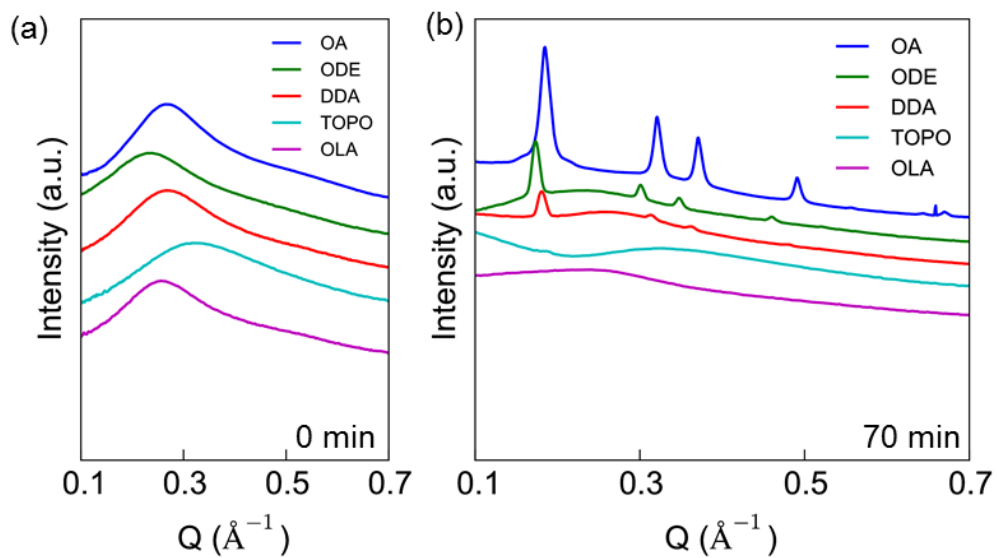

**Figure S26**: MSC reaction with *1000 mM Cd oleate in various solvent—precursor and mesophase structure comparison (SAXS/WAXS).* (a) Cd oleate precursor structure at 0 min at 130 °C. The OA is the original 1000 mM reaction mixture (with free oleic acid, OA). The Cd oleate micelle peak shift to larger d-spacing (smaller Q) for the ODE and OLA samples, and to smaller d-spacing (large Q) for TOP=O sample. The DDA sample d-spacing matches the OA sample. (b) The SAXS pattern at 70 min at 130 °C. The OA, ODE, DDA samples show a hexagonal mesophase. The TOP=O sample shows a small peak around the location of the 1Q mesophase peak as well as a steeper slope at low Q (<0.2 Å$^{-1}$) indicative of larger nanoparticles. The OLA sample does not show a mesophase peak, but does show a change in the low Q structure factor (0.1-0.2 Å$^{-1}$) compared to precursor structure (a), likely indicating MSC or NP formation.

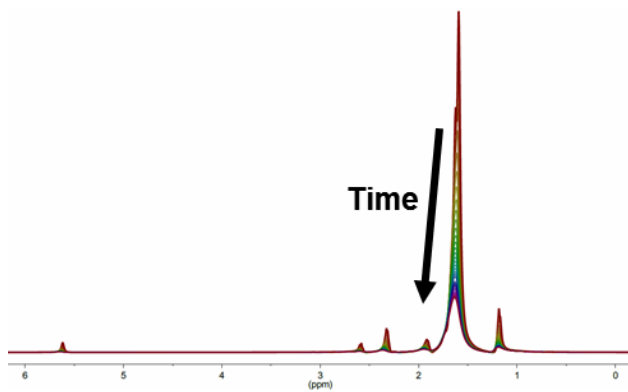

**Figure S27**: *In-situ $^1$H NMR at 130 °C for 1000 mM reaction.* **Figure 8a** shows that the total spectral area decreases with time.

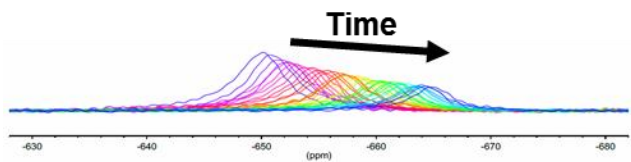

**Figure S28**: *In-situ $^{113}$Cd NMR at 130 °C for 1000 mM reaction.* **Figure 8a** shows that the total spectral area decreases with time.

Kinetics Fits

**Table S5**: *Fitting parameters and $R^2$ values for the NMR silencing rate (see **Figure 8a**).*

| Species | A | k (s$^{-1}$) x 10$^4$ | R$^2$ |
|---|---|---|---|
| H1[a] | 0.68 | 0.91 | 0.991 |
| C13[a] | 0.56 | 0.95 | 0.982 |
| P31[a] | 0.83 | 1.69 | 0.999 |
| Cd113[a] | 0.35 | 0.16 | 0.775 |

[a]Fit: $1 - A \cdot (1 - \exp(-kt))$

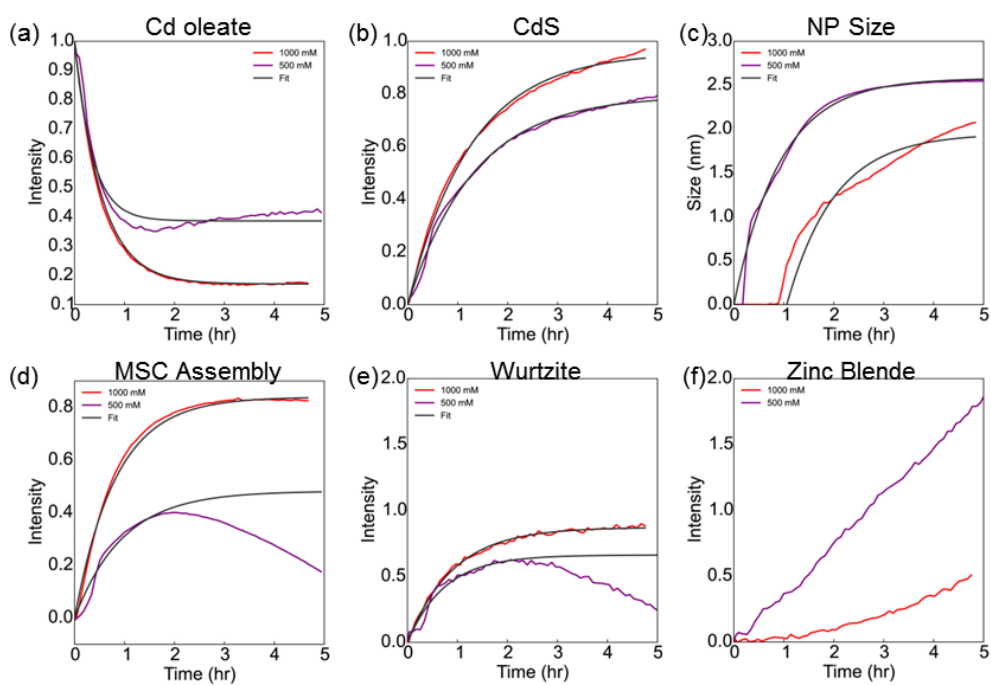

**Figure S29**: *Fits to SAXS/WAXS peak evolutions.* (a) Cd oleate, (b) CdS formation (1.87 Å$^{-1}$), (d) MSC assembly (first mesophase peak), and (e) the WZ contribution to the diffraction are fits to data shown in **Figure 6**. The rates are roughly $10^{-4}$ s$^{-1}$. (c) Fit of growth rate based on size determined from the monodisperse spherical form factor fits (**Figure S11**). (f) the ZB contribution to the diffraction was not fit because the trend is linear, and the intensity is not quantitative.

**Table S6**: Fitting parameters for X-ray data in **Figure S29**.

| | A | k (1/s) x $10^4$ | $R^2$ |
|---|---|---|---|
| Cd oleate 1000 mM[a] | 0.83 | 5.2 | 0.998 |
| Cd oleate 500 mM[a] | 0.61 | 7.9 | 0.958 |
| MSCs 1000 mM[b] | 0.83 | 3.4 | 0.995 |
| MSCs 500 mM[b,c] | 0.48 | 2.9 | 0.970 |
| CdS 1000 mM[b] | 0.96 | 2.2 | 0.997 |
| CdS 500 mM[b] | 0.79 | 2.1 | 0.996 |
| NP 1000 mM[b,d,e] | 1.95 | 0.049 | 0.853 |
| NP 500 mM[b,d] | 2.58 | 0.050 | 0.989 |
| WZ 1000 mM[b] | 0.87 | 3.1 | 0.996 |
| WZ 500 mM[b,c] | 0.66 | 3.7 | 0.965 |

[a]Fit: $1 - A \cdot (1 - \exp(-kt))$. [b]Fit: $A \cdot (1 - \exp(-kt))$.
[c]Fit from 0 to 2. The intensity linearly decreases after that, but is not calibrated to quantitative. [d]Rate of growth, not rate of formation. [e]Fit start at 1 h.

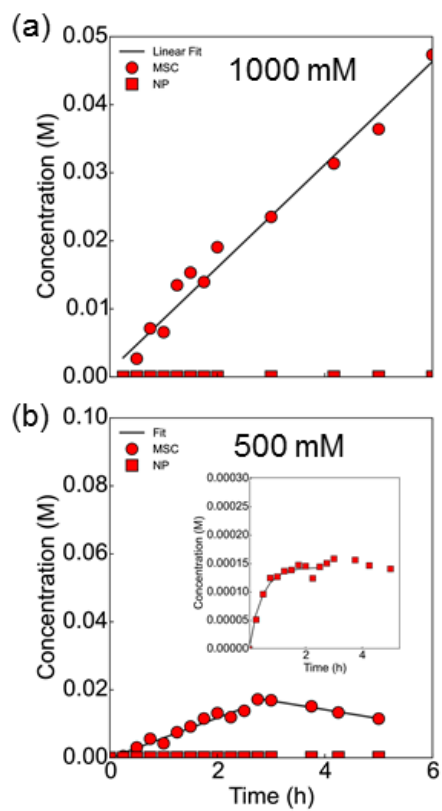

**Figure S30**: *MSC formation rates (ex situ absorbance).* The linear fits (zeroth order rates) of the MSC formation (**Figure S1b,S1e**). (a) For the 1000 mM, the MSC formation rate constant is 2.1 x $10^{-6}$ M/s ($R^2$=0.977). The rate of formation of NPs is near zero. (b) For 500 mM, the MSC formation rate at early times (<2.7 h) is 1.6 x $10^{-6}$ M/s ($R^2$ = 0.956) and at later times, the MSC depletion rate is -7.3 x $10^{-7}$ M/s ($R^2$ = 0.991). Inset: In 500 mM reaction, the NP formation rate is first order (6 x $10^{-4}$ 1/s, $R^2$ = 0.974) and plateaus at longer time (>2 h).

# Calculations

## Purity Calculation

The purity was calculated based on the concentration of the particles, using a published size and size-dependent extinction coefficient curves for cadmium sulfide.[9] The size curve uses the wavelength in nm to calculate the particle diameter in nm.

$$d = -6.6521 \times 10^{-8} \cdot \lambda^3 + 1.9557 \times 10^{-4} \cdot \lambda^2 - 9.2352 \times 10^{-2} \cdot \lambda + 13.29$$

The size-dependent extinction coefficient ($\epsilon$) is calculated based on the diameter in nm, and provides the extinction coefficient in units of cm$^{-1}$ M$^{-1}$.

$$\epsilon = 21536 \cdot d^{2.3}$$

The concentration was determined using the empirical extinction coefficient and Beer's law:

$$C = \frac{A}{\epsilon L},$$

where A is the absorbance at the wavelength correspond to particle size, and L is the path length (1 cm).

The selectivity was calculated based on the MSC and NP concentrations as follows:

$$Selectivity = \frac{C_{MSC}}{C_{NP}}.$$

The purity was calculated based on the MSC and NP concentrations as follows:

$$Purity = \frac{C_{MSC}}{C_{MSC} + C_{NP}} \cdot 100.$$

## MSC Conversion

The conversion of MSC compared to the maximum yield was determined as follows:

$$Conversion = \frac{Mass\ of\ produced\ MSCs}{Theoretical\ mass\ of\ MSCs} \cdot 100.$$

The mass produced of MSCs was calculated as follows:

$$Mass\ produced\ MSC = A_{MSC} \cdot CF \cdot DF \cdot wt_{inorg} \cdot Vol_{flask},$$

where $A_{MSC}$ is the absorbance at 324 nm (0.79 at 6 h for 1000 mM reaction), CF is the calibration factor from MSC concentration per absorbance based on a known concentration of particles (CF=0.0526 g solid product/(L· $A_{MSC}$)), DF is the dilution factor from flask concentration (4000), $wt_{inorg}$ is the weight percent of the inorganic component based on ICP (29.1% from ref.[1]) and the flask volume is 23.5 mL.

The theoretical mass of MSCs was calculated as follows based on:

$$Theoretical\ mass\ of\ MSCs = \frac{mass_{CdO}}{Mw_{CdO}} \cdot v_{Cd:S} \cdot Mw_{MSC_{inorg}}$$

where $Mw_{CdO}$ and $Mw_{MSC_{inorg}}$ are the molecular weights of cadmium oxide and inorganic (cadmium sulfide component with 2:1 ratio based on ref.[1]) for the MSCs, respectively; and $v_{Cd:S}$ is the composition of the clusters 2:1 Cd:S ratio.[1]

For instance, the conversion and theoretical mass are

$$Theoretical\ mass\ of\ MSCs = \frac{2.56\ g}{128.413\ \frac{g}{mol}} \cdot \frac{1}{2} \cdot 256.88\ \frac{g}{mol} = 2.56\ g\ MSCs\ (inorganic).$$

$$Conversion = \frac{0.790 \cdot 0.0526\ \frac{g}{L} \cdot 4000 \cdot 0.291 \cdot 0.0235L}{2.56\ g\ MSCs\ (inorganic)} \cdot 100 = 44\%$$

### Size dispersion

Based on the FHWM in absorbance, the size dispersion based on Peng curve (above) is as follows (the 2.355 factor converts from FWHM to standard deviation:

$$\sigma_{Abs} = \frac{(328\ nm) - (320nm)}{2.355},$$

$$\sigma_d = d(324\ nm + \sigma_{Abs}) - d(324nm),$$

$$\sigma_r = \frac{\sigma}{\mu}.$$

For instance, for the 324 nm (8 nm FWHM) MSC the size dispersion is the following:

$$\sigma_{Abs} = \frac{(328\ nm) - (320nm)}{2.355} = 3.4\ nm$$

$$\sigma_d = d(324\ nm + 3.4\ nm) - d(324nm) = 0.047\ nm, and$$

$$\sigma_r = \frac{0.047\ nm}{1.636\ nm} = 2.9\%.$$

Notably, the size standard deviation of MSC is ~ 0.5 Å (which is less than 20% of cadmium sulfide bond length, ~2.5 Å, and gives 3% size dispersion, which is narrower that best NP samples to date.[12–14]

### Instrumental Line Broadening

To calculate the line broadening (q), the following analysis was used. By differentiation and substituting for q, the following propagation of error expression is obtained:

$$q = 2\pi k sin(\theta) = \frac{4\pi sin(\theta)}{\lambda}$$

$$dq = \sqrt{\left(-\frac{4\pi \sin(\theta)}{\lambda^2} d\lambda\right)^2 + \left(\frac{4\pi}{\lambda \tan(\theta)}\right)^2}$$

$$\frac{dq}{q} = \sqrt{\left(\frac{dk}{k}\right)^2 + \left(\frac{d\theta}{\tan(\theta)}\right)^2}$$

$$\frac{dk}{k} = \frac{d\lambda}{\lambda} = \frac{dE}{E} = \frac{20eV}{20keV} = 10^{-3}$$

$$dth = \frac{source\ size}{distance_{source-to-slit}} = 1.2x10^{-4}$$

For instance, at first hexagonal peak (q=0.1845 Å$^{-1}$), the instrumental line broadening is

$$\frac{dq}{q} = \sqrt{(10^{-3})^2 + \left(\frac{1.2x10^{-4}}{\tan(\theta)}\right)^2}$$

For instance, dq$_{int}$ for instrument at the first SAXS peak is

$$dq_{int} = \sqrt{(10^{-3})^2 + \left(\frac{1.2x10^{-4}}{\tan\left(\frac{0.0182}{2}\right)}\right)^2}\ 0.1845\text{Å}^{-1} = 0.00244\ \text{Å}^{-1}.$$

Next, the peak position and wavelength values need to be converted from q to theta using the following equation:

$$\theta = asin\left(\frac{q\lambda}{4\pi}\right), and$$

the corresponding propagation of error equation by differentiation as

$$\frac{d}{dx} asin(ax) = \frac{a}{\sqrt{1-a^2x^2}},$$

leading to

$$d\theta = \sqrt{\left(\frac{\frac{q}{4\pi}}{\sqrt{1-\frac{q}{4\pi}^2 \lambda^2}}\right)^2 (d\lambda)^2 + \left(\frac{\frac{\lambda}{4\pi}}{\sqrt{1-\frac{\lambda}{4\pi}^2 q^2}}\right)^2 (dq_{int})^2},$$

where $\lambda$ is the beam wavelength (0.62054 Å).

For instance, the angle and error are

$$\theta = asin\left(\frac{0.1845\text{Å}^{-1} \cdot 0.62054\text{Å}}{4\pi}\right) = 0.00911, and$$

$$d2\theta_{FWHM} = 2 \cdot 2.355 \sqrt{\left(\frac{\frac{0.1845\text{Å}^{-1}}{4\pi}}{\sqrt{1-\left(\frac{0.1845\text{Å}^{-1}}{4\pi}\right)^2 (0.62054\text{Å})^2}}\right)^2 (10^{-3} \cdot 0.62054\text{Å})^2 + \left(\frac{\frac{0.62054\text{Å}}{4\pi}}{\sqrt{1-\left(\frac{0.62054\text{Å}}{4\pi}\right)^2 (0.1845\text{Å}^{-1})^2}}\right)^2 (0.00244 \text{ Å}^{-1})^2} = 5.68 \cdot 10^{-4}$$

where the 2.355 is to convert from standard deviation to FWHM.

### Scherrer Calculation

The crystallite size is calculated with the Scherrer equation:

$$d_{crystal} = \frac{K\lambda}{FWHM \cos(\theta)}$$

where K is a shape factor constant (0.9), FWHM is the FWHM for experimental peak subtracted from the FWHM for the instrument in quadrature, and theta is the peak position. FWHM is calculated as follows:

$$FWHM = \sqrt{(\Delta 2\theta_{exp})^2 - (\Delta 2\theta_{FWHM})^2},$$

$$FWHM = \sqrt{(6.568 \cdot 10^{-4})^2 - (5.68 \cdot 10^{-4})^2} = 3.29 \cdot 10^{-4}$$

For instance, the broadening (or crystallize size) for the first SAXS peak is

$$d_{crystal} = \frac{0.9 \cdot 0.62054 \text{ Å}}{(3.29) \cdot 10^{-4} \cdot \cos(0.00911)} = 170 \ nm$$

Note that the percentage of the instrumental broadening that contributes to the experimental broadening is 86% (=instrumental over experimental broadening or 5.68/6.568), indicating that the peak width is near the instrumental limit as well as the applicable limit of the Scherrer equation.